\documentclass{ws-mpla}
\usepackage[super]{cite}
\usepackage{graphicx}
\usepackage{color}
\usepackage{subfigure,float, afterpage}  
\usepackage{hyperref}
\newcommand{\beq}{\begin{equation}}
\newcommand{\eeq}{\end{equation}}
\newcommand{\bea}{\begin{eqnarray}}
\newcommand{\eea}{\end{eqnarray}}

\def\D21{\Delta m_{21}^2}
\def\D32{\Delta m_{32}^2}
\def\sij{\sin\Theta_{ij}}
\def\cij{\cos\Theta_{ij}}
\def\s12{\sin\Theta_{12}}
\def\s23{\sin\Theta_{23}}
\def\s13{\sin\theta_{13}}

\def\t12{\Theta_{12}}
\def\t23{\Theta_{23}}
\def\t13{\Theta_{13}}

\def\l{\left}
\def\r{\right}

\def\la{\mathrel{\mathchoice {\vcenter{\offinterlineskip\halign{\hfil
$\displaystyle##$\hfil\cr<\cr\sim\cr}}}
{\vcenter{\offinterlineskip\halign{\hfil$\textstyle##$\hfil\cr<\cr\sim\cr}}}
{\vcenter{\offinterlineskip\halign{\hfil$\scriptstyle##$\hfil\cr<\cr\sim\cr}}}
{\vcenter{\offinterlineskip\halign{\hfil$\scriptscriptstyle##$\hfil\cr<\cr\sim
\cr}}}}}

\begin{document}

\title{Velocity Induced Neutrino Oscillation and its Possible 
Implications for Long Baseline Neutrinos}

\author{\footnotesize AMIT DUTTA BANIK\footnote{
email: amit.duttabanik@saha.ac.in}} 

\author{\footnotesize DEBASISH MAJUMDAR\footnote{
email: debasish.majumdar@saha.ac.in}}

\address{Astroparticle Physics and Cosmology Division, \\
Saha Institute of Nuclear Physics \\
1/AF Bidhannagar, Kolkata 700064, India.}

\maketitle

\begin{abstract}
If the three types of active neutrinos possess different maximum 
attainable velocities and the neutrino eigenstates in the velocity 
basis are different from those in the flavour (and mass) basis then 
this will induce a flavour oscillation in addition to the normal 
mass flavour oscillation. Here we study such an oscillation scenario 
in three neutrino framework including also the matter effect and apply 
our results to demonstrate its consequences for long baseline neutrinos. 
We also predict the possible signatures in terms of yields in a possible 
long baseline neutrino experiment. 

\end{abstract}

\ccode{PACS Nos.: 14.60.Pq}

\section{Introduction}

It is now established that neutrinos exhibit the phenomenon of oscillation 
whereby one type of neutrino (electron, muon or tau) can change to another
flavour when they propagate through vacuum or matter. 
It is also established that they are massive and the fact that their 
mass eigenstates and their weak interaction eigenstates are not 
same, leads to the phenomenon of oscillation. The origin of mass for the
neutrinos cannot be explained within the framework of Standard Model of 
particle physics and one may need theories beyond Standard
Model (SM) to explain how neutrinos acquire mass. Thus neutrinos 
provide us an window for new physics beyond SM.
Moreover, neutrino can play a key role in addressing the 
CP violation in leptonic sector. The neutrinos, if  Majorana
fermions, could be a useful probe to address the problem 
of leptogenesis and explain the phenomenon of neutrinoless double beta decay.  

The oscillation in neutrinos can also be introduced in other 
exotic scenarios such as violation of equivalence principle (VEP)
whereby three types of neutrinos interact gravitationally with different 
strengths and thus their eigenstates in gravity basis is not the same as 
those in interaction basis. This scenario has been explored in some previous 
works \cite {smirnov, utpal, arunansu}. Neutrino flavour oscillations,
other than mass-flavour oscillation, can also 
be initiated if three different neutrinos have different maximum attainable
velocities. This may be possible when the three massive neutrinos
have different masses.
In this scenario the mixing between the flavour 
and velocity eigenstates can induce neutrino flavour oscillation.
It is to be noted that while the oscillation length $\lambda \sim E_{\nu}$
($E_{\nu}$ is the neutrino energy) for ordinary mass-flavour oscillation,
for the cases of both velocity induced flavour oscillation and  
VEP induced oscillation, $\lambda \sim \frac {1} {E_{\nu}}$. In fact, 
for velocity induced oscillation, $\lambda = \frac {2\pi} {E_{\nu} \Delta V}$
where $\Delta V$ is the difference in maximum attainable velocities 
of two neutrinos. In case of mass-flavour oscillation 
$\lambda = \frac {4\pi E_{\nu}} {\Delta m^2}$, where $\Delta m^2$ is the 
mass square difference between two neutrinos. 

Studies of velocity induced neutrino oscillation have been performed by
previous authors \cite {coleman},
where they consider the violation of Lorentz invariance whereby 
one of the neutrinos may acquire velocity higher than $c$, the velocity
of light. 
Detailed study on the prospect of
velocity induced oscillation in two flavour scenario is given 
in Coleman and Glashow\cite{coleman}. Such study, with experimental data
considering $\nu_\mu \rightarrow \nu_e$ maximal mixing, predicts that   
$\Delta V < 10^{-21}$. 
There are other works too, where this issue has been addressed.
Fogli {\it et al}\cite{fogli} has made an analysis of atmospheric
neutrino data from Super-Kamiokande experiment for both the 
velocity induced and VEP induced oscillations and
obtained a bound $\Delta V < 10^{-24}$.
In a later work by Battistoni {\it et al} \cite{battistoni},
the data of upward going muons for atmospheric neutrinos in
MACRO experiment\cite{macro,ambrosio} has been analysed considering both
mass flavour oscillation and velocity induced oscillation.
This analysis gave a bound (with 90\% C.L.) as $\Delta V < 6 \times
10^{-24}$ for the velocity-flavour state mixing angle $\theta_v =0$
and $\Delta V < (2.5 - 5) \times 10^{-26}$ for $\theta_v = \frac {\pi} {4}$.
But in both the analyses above, only
2-flavour oscillation, $\nu_\mu \rightarrow \nu_\tau$, is considered.
The matter effect on neutrino oscillation is irrelevant for only
$\nu_\mu \rightarrow \nu_\tau$ oscillation scenario (since the matter
effect term is same for both $\nu_\mu$ and $\nu_\tau$, no phase
difference can be generated). Thus both analyses are for vacuum oscillation
only. In a more recent work \cite{gonzalezgarcia}, it is shown from the
analysis of atmospheric and long baseline neutrino (LBL) data that the
bound on $\Delta V$ for $\nu_\mu \rightarrow \nu_\tau$ oscillation with
CPT even effects is $\leq 10^{-24}$.

In this work we consider three types of neutrinos having different 
velocities such that their velocity eigenstates differ
from their flavour eigenstates and mass eigenstates.
We investigate the combined oscillation phenomenology for both mass-flavour and
velocity flavour oscillations.
We explore how the velocity induced oscillation changes when the 
matter effect is included. As discussed earlier,
in this scenario, the oscillation depends both on the 
difference of velocities $\Delta V$ of any two neutrinos.   
We investigate the possible magnitudes for $\Delta V$ in order 
that this velocity-driven oscillation (with $\Delta m^2$) can have
significant effects
in the combined oscillation scenario (mass induced and velocity induced 
oscillation with matter effect) mentioned above.
Our formalism is then applied to demonstrate the nature of this velocity
induced oscillation for three massive neutrinos with and without the matter
effect for a chosen baseline through earth matter. We then compute the number of 
neutrino induced muons for such a scenario in case of a possible long baseline
experiment with an iron calorimeter (ICAL) as the end detector.

The paper is organised as follows. In Section 2 we give the 
general framework for neutrino oscillation formalism of massive induced by 
the velocity eigenstates of neutrinos including the matter
effect, as also for the case where matter effect is not included.
The possible signature of such velocity induced oscillation in a long 
baseline experiment is probed considering a specific example for 
demonstrative purpose. This is given in Sect. 3.
Finally in Sect. 4 we make some concluding remarks.

\section{Neutrino Oscillation Framework with Velocity Eigenstates}
The basic idea of neutrino oscillation, first given by 
Pontecorvo \cite{pontecorvo,pontecorvo1},
comes from simple quantum mechanics of a 
two - level quantum system. Let the system be in one of its 
stationary states. Let an eigenstate of its Hamiltonian be expressed 
as $| \psi_i \rangle$ whose time evolution 
is  written as 
$| \psi _{i} (t) \rangle = e^{-iE_{i}t}| \psi_{i} (0)  \rangle$. 
If a state is produced which is not one of the eigenstates of the 
Hamiltonian, the probability that 
the system retains its state will then oscillate with time 
depending on the frequency $\omega _{21} = E_{2} - E_{1}$, where 
$E_{1}$ and $E_{2}$  are the energy eigenvalues
of the two level quantum system. 
Since the eigenstates in the mass (energy) basis of a neutrino 
are not the same as those in interaction or flavour basis, 
a neutrino produced in certain flavour eigenstate $|\nu_\alpha \rangle$
of flavour $\alpha$ can oscillate into 
other flavour with a definite probability or
come back to its initial flavour state as well.             
The neutrino flavour eigenstates and mass eigenstates are related through 
a unitary matrix $U$  which can be parametrised as 
\bea
\left( \begin{array}{c}
               |\nu_e \rangle\\ 
               |\nu_\mu\rangle\\ 
               |\nu_\tau\rangle 
              \end{array} \right) &=& 
\left(\begin{array}{ccc} U_{e1} & U_{e2}& U_{e3} \\
                        U_{\mu1}& U_{\mu2}& U_{\mu3}  \\
                        U _{\tau1}& U_{\tau2}& U_{\tau3}   \end{array}\right)
\left(\begin{array}{c}|\nu_1\rangle\\
                      |\nu_2\rangle  \\
                      |\nu_3\rangle  \end{array} \right). 
\label{eq1}
\eea
Therefore, a flavour eigenstate $|\nu_\alpha \rangle$ ($\alpha = e,\mu\, 
{\rm or}\, \tau$) is related 
to mass eigenstates $|\nu_i \rangle (i=1,2,3)$ by the relation 
\bea
| \nu_{\alpha} \rangle &=& \sum_i U_{\alpha i} | \nu_{i} \rangle\,\, .
\label{eq2}
\eea
Assuming the neutrinos to be CP conserving,
the mixing matrix $U$ can be expressed 
in terms of three orthogonal matrices, $V_{23}$, $V_{13}$ and $V_{12}$ 
(three successive rotations) as \cite{giunti,akhmedov}
\bea
U &=& V_{23} V_{13} V_{12} \nonumber \\
&=& \left ( \begin{array}{ccc} 1 &   0    &   0 \\
                               0 & c_{23} & s_{23} \\
                               0 & -s_{23} & c_{23} \end{array} \right )
\left ( \begin{array}{ccc} c_{23} & 0 & s_{13} \\
                             0    & 1 &   0    \\
                          -s_{23} & 0 & c_{23} \end{array} \right )
\left ( \begin{array}{ccc} c_{12} & s_{12} & 0 \\
                          -s_{12} & c_{12} & 0 \\
                             0    &   0    & 1 \end{array} \right ). 
\label{eq3}
\eea
The mixing matrix $U$ takes the form
\bea
  U &=&\left( \begin{array}{ccc}
    c_{13}c_{12}                    & s_{12}c_{13} & s_{13}      \\
   -s_{12}c_{23}-s_{23}s_{13}c_{12} &   c_{23}c_{12}-s_{23}s_{13}s_{12} 
                                                  & s_{23}c_{13}  \\
    s_{23}s_{12}-s_{13}c_{23}c_{12} & -s_{23}c_{12}-s_{13}s_{12}c_{23}   
                                                  & c_{23}c_{13}       
                                 \end{array}  \right)\,\, .
\label{eq4}
\eea   
In Eqs. (\ref {eq3}) - (\ref {eq4}), $c_{ij} \equiv \cos\Theta_{ij}$ 
and $s_{ij} \equiv \sin\Theta_{ij}$
and $\Theta_{ij}$ is the flavour mixing angle between $i^{\rm th}$ and 
$j^{\rm th}$ neutrinos having mass eigenstate $|\nu_i\rangle$  
and $|\nu_j\rangle$ respectively. Needless to mention 
that $U$ is a unitary matrix. 

The time evolution of the neutrinos with mass eigenstate 
$|\nu_i \rangle,\, (i=1,2,3)$ is (in natural units) given by 
\bea
i\frac {d} {dt} \left ( \begin{array}{c} |\nu_1 \rangle \\
                                          |\nu_2 \rangle \\
                                          |\nu_3 \rangle \end{array} \right )
&=& \l ( \begin{array}{ccc}
                        E_1 & 0 & 0 \\
                        0  & E_2 & 0 \\
                        0  & 0 & E_3   \end{array} \r )
\left ( \begin{array}{c} |\nu_1 \rangle \\
                     |\nu_2 \rangle \\
                     |\nu_3 \rangle \end{array} \right )
= H_d \left ( \begin{array}{ccc} |\nu_1 \rangle \\
                                 |\nu_2 \rangle \\
                                |\nu_3 \rangle \end{array} \right ) \nonumber 
\eea
Using Eqs.~(\ref{eq1}), (\ref{eq2}), one gets the evolution equation 
in flavour basis as 
\bea
i\frac {d} {dt} U^{\dagger} \left ( \begin{array}{c} |\nu_e \rangle \\
                                          |\nu_\mu \rangle \\
                                          |\nu_\tau \rangle \end{array} \right )
&=& H_d U^{\dagger}\left ( \begin{array}{c} |\nu_e \rangle \\
                                 |\nu_\mu \rangle \\
                                 |\nu_\tau \rangle \end{array} \right ), \nonumber 
\eea
\bea
i\frac {d} {dt} \left ( \begin{array}{c} |\nu_e \rangle \\
                                          |\nu_\mu \rangle \\
                                          |\nu_\tau \rangle \end{array} \right )
&=& H \left ( \begin{array}{c} |\nu_e \rangle \\
                               |\nu_\mu \rangle \\
                               |\nu_\tau \rangle \end{array} \right )
\label{eq5}
\eea
where
\bea
H &=& U H_d U^{\dagger} 
\label{eq6}
\eea
In the above, $H_d$ is the Hamiltonian in mass basis with the energy
and mass eigenvalues of the three neutrinos are denoted 
as $E_i,~i = 1,2,3$ and $m_i,~i=1,2,3$. Now the diagonal matrix $H_d$
can be written in the form  
\bea
H_d &=& \l ( \begin{array}{ccc} 
                        E_1 & 0 & 0 \\
                        0  & E_2 & 0 \\ 
                        0  & 0 & E_3   \end{array} \r ) 
\approx \frac {1} {2E} \l ( \begin{array}{ccc}
                        m_1^2 & 0 & 0 \\
                        0  & m_2^2 & 0 \\
                        0  & 0 & m^2_3   \end{array} \r ) \\
    &\approx& \frac{1}{2E}{\rm diag}(-\Delta m_{21}^2,0,\Delta m_{32}^2)\, .
\label{eq8}
\eea 
Eq.~(7) is obtained with the approximation that the 
neutrino momenta $p_i \approx E_i$ 
($i=1,2,3$) 
and $p_1 \approx p_2 \approx p_3 =p \approx E$. Then,   
$E_i = \sqrt {p_i^2 + m_i^2} \approx p + \frac {m_i^2} {2p} \approx 
p + \frac {m_i^2} {2E}$.
In Eq.~(\ref{eq8}),
$\Delta m_{ij}^2$ denotes the difference of the square
of the masses of $i^{\rm th}$ and  $j^{\rm th}$ neutrinos. The matrix  
$\frac {1} {2E} {\rm diag}(m_2^2,m_2^2,m_2^2)$ which is subtracted from
Eq.~(7) to obtain Eq.~(\ref{eq8}) as also the matrix 
${\rm diag}({p,p,p})$ which are neglected in Eq.~(7)
induce no phase difference between 
any two neutrinos and hence do not contribute to oscillation.  
The flavour oscillation probability (mass-induced) for a neutrino of flavour 
$\alpha$ ($|\nu_\alpha \rangle$, $\alpha$ denotes $e$ or $\mu$ or $\tau$)
after a time $t$ when it traverses a distance $L = ct$, can 
now be obtained by solving Eq.~(\ref{eq5}) and calculating 
the quantity $|\langle \nu_\alpha (t) | \nu_\alpha (t=0) \rangle |^2$
(the square of the ampliude). We mention in the passing that oscillation 
probability for pure 
mass-flavour vacuum oscillation considering only two neutrino species 
is given by
\bea
P_{12}=\sin^2\theta \sin^2\left(\frac{\Delta m^2_{21} L}{4E}\right), \nonumber
\eea
where $\theta$ is the mixing angle between the two active neutrino species. 
For the case of mass-flavour oscillation, the oscillation probability
depends on the phase difference which is $\sim \frac{1}{E}$ (oscillation length 
$\lambda = \frac{4\pi E}{\Delta m^2_{21}}$) and this also holds for the 
case of oscillation of three active neutrinos.

Massive neutrinos may also exhibit oscillation 
phenomena if different 
neutrinos have different maximum attainable 
velocities and they travel at different speeds in vacuum. 
As they move with different velocities they must differ in energy.
If the velocity eigenstates are not the same as flavour
eigenstates then similar to the mass induced oscillation, one has  
velocity-flavour oscillation induced by different maximum velocities 
of neutrinos.
In such a scenario, the velocity eigenstate 
$\left |\nu_{V_i}\right \rangle,~i=1,2,3$ is related to 
the flavour eigenstate $|\nu_\alpha \rangle, ~\alpha \equiv e,\mu,\tau$
through three mixing angles $\Theta^{'}_{ij} (i \neq j),~i,j = 1,2,3$.
The difference in velocities between any two species of neutrinos 
moving with different maximum attainable velocities can 
be expressed as
\bea
\Delta V_{ji} &=& V_j - V_i = \frac {p_j} {E_j} - \frac {p_i} {E_i} \,, 
\nonumber \\
\Delta V_{ji} E &=& p_j - p_i~; (E_j \approx E_i \approx E) , \nonumber
\eea
where $V_{i({\rm or}~j)}$, $p_{i({\rm or}~j)}$, $E_{i({\rm or}~j)}$ 
respectively denote the maximum attainable velocity, momentum and energy
for the neutrino species $|\nu_{i({\rm or}~j)}\rangle$.
With the approximation that $E_j = p_j$ and $E_i = p_i$, in the above equation
the energy difference between any two species of neutrinos $|\nu_j \rangle$
and $|\nu_i \rangle$ is obtained as\cite{coleman}
\bea 
p_j - p_i \approx E_j - E_i = \Delta E_{ji} &=& \Delta V_{ji} E. \nonumber
\eea
For a pure velocity induced oscillation of two active neutrinos with mixing
angle $\theta_v$, the oscillation probability is of the form
\bea
P'_{12}=\sin^2\theta_v \sin^2\left(\frac{\Delta V_{21} L E}{2}\right). \nonumber
\eea
In the case of pure velocity induced oscillation, the phase difference
is $\sim E$ ($\lambda = \frac{2\pi}{E \Delta V_{21}}$).
Massive neutrinos may undergo simultaneous mass and velocity induced 
oscillations. In the presence of both mass and velocity mixings of three 
neutrino species, the 
effective Hamiltonian of the system in flavour basis can be 
constructed as 
\beq
H'= UH_dU^\dagger +U^{'}H_{d}^{'}U^{'\dagger} 
\label{eq9}
\eeq 
where $H_d$ is given in Eq. (\ref{eq8}) and
\beq 
H_{d}^{'} = E~{\rm diag} (\Delta V_{21},0,\Delta V_{32})\, .
\label{eq10}
\eeq
The flavour-velocity mixing matrix $U^{'}$ for 3-neutrino species
can also be written in the 
similar form of Eq.~\ref{eq4} as
\bea
U^{'} &=&  \left( \begin{array}{ccc}
         c_{13}'c_{12}' & s_{12}'c_{13}' & s_{13}'   \\
-s_{12}'c_{23}'-s_{23}'s_{13}'c_{12}' & c_{23}'c_{12}'-s_{23}'s_{13}'s_{12}'
                                      & s_{23}'c_{13}'  \\
 s_{23}'s_{12}'-s_{13}'c_{23}'c_{12}' & -s_{23}'c_{12}'-s_{13}'s_{12}'c_{23}'
                                      &  c_{23}'c_{13}'       
                                 \end{array}  \right) 
\label{eq11}
\eea
where $s_{ij}' = \sij'$ and $c_{ij}' = \cij'$ and $\Theta'_{ij}$ is the
mixing angles relating neutrinos in flavour basis to that of velocity 
basis. 

It is well known that the neutrino oscillation is modified 
when neutrinos pass through matter. The Mikheyev, Smirnov, Wolfenstein
(MSW) effect or
matter effect on neutrino oscillation is caused by the interaction
of neutrinos with matter.
The matter effect can be included in the oscillation formalism by 
modifying the Hamiltonian in Eq.~(\ref{eq9}) as
\bea  
H'' &=& UH_dU^\dagger +U^{'}H_{d}^{'}U^{'\dagger} + V_{cc}
\label{eq12}
\eea
where $V_{cc}$ denotes the potential (matter potential) responsible for the 
interaction of neutrinos with matter during their passage through
matter. The matter potential $V_{cc}$ can be calculated as 
\bea
V_{cc}&=&{\rm diag}(\sqrt{2}G_FN_e,0,0)
\label{eq13}
\eea
where $N_e$ is the electron number density and $G_F$ is the Fermi constant. 
It may be noted that for 
antineutrinos, the interaction potential $V_{cc}$ is replaced by $-V_{cc}$.  
With matter effect, the evolution equation in Eq.~(\ref{eq5}) will 
now be modified as
\bea
i\frac{d}{dt}|\nu_\alpha\rangle &=& H'' |\nu_\alpha\rangle .
\label{eq14}
\eea          

Now, for the purpose of demonstration of the effect of velocity 
induced oscillation
in presence of matter and for the simplicity of the calculation, 
we assume a special case where
mass mixing angles and velocity mixing angles with flavour states 
are same. Then we have $U = U'$ in our formalism and the evolution equation 
for a neutrino of flavour $\alpha$ takes a simplified form
\bea
i\frac{d}{dt}|\nu_\alpha\rangle &=& H_{f} |\nu_\alpha\rangle\, , 
\label{eq15}
\eea
where
\bea
H_{f} &=& U(H_d +H_{d}^{'})U^{\dagger}+V_{cc}\, . 
\label{eq16}
\eea 
Thus the Hamiltonian $H_f$ is written as 
\bea
H_{f} &=& U~{\rm diag} (-\frac{\Delta m_{21}^{2}}{2E} - 
\Delta V_{21}E,~0, ~\frac{\Delta m_{32}^{2}}{2E}+\Delta V_{32}E)~U^{\dagger} + 
V_{cc}\nonumber \\
       &=& \frac{1}{2E} U~ {\rm diag}(-\Delta m_{21}^{'2},0,
\Delta m_{32}^{'2})~ U^{\dagger}+ V_{cc}\, .
\label{eq17}
\eea
In the above
\bea
\frac{\Delta m_{21}^{'2}}{2E} &=& \frac{\Delta m_{21}^{2}}{2E} + \Delta V_{21} 
E \, , \nonumber \\
\frac{\Delta m_{32}^{'2}}{2E} &=& \frac{\Delta m_{32}^{2}}{2E} + \Delta V_{32} 
E \, . 
\label{eq18}
\eea
We focus our discussions for 
the cases where $|\Delta m_{21}^{'2}| << |\Delta m_{32}^{'2}|$ such that
$|\Delta m_{21}^{'2}|$  can be neglected. 
The mass square difference, $\Delta m_{32}^2 (\sim 10^{-3} {\rm eV}^2)$ 
is experimentally found 
to be larger than $\Delta m_{21}^2 (\sim 10^{-5} {\rm eV}^2)$ and 
the former is relevant for 
atmospheric neutrinos, long baseline neutrinos
etc. where the oscillation is effective for the baseline 
lengths $\sim 10^{3} - 10^4$ Km ($\sim$ earth's diameter).
For such a scenario the velocities of neutrinos in the present case 
are so assumed that  
$|\Delta V_{21}| << |\Delta V_{32}|$ and 
$|\Delta m_{21}^{'2}| << |\Delta m_{32}^{'2}|$ from Eq.~(\ref{eq18}) and
oscillation length $L_{\rm osc} = \frac{4\pi E}{\Delta m_{32}^{'2}}$ 
is comparable with the baseline lengths relevant for atmospheric
or long baseline neutrinos. 
With these assumptions, 
the probabilities of oscillations $P_{\alpha_1 \alpha_2}$ 
from one flavour $\alpha_1$ to 
another flavour $\alpha_2$ for constant matter density 
takes the form \cite{gonzalez}
\bea
        P_{ee}&=&1-4s_{13,m}^{2} c_{13,m}^ {2}S^2_{31}\, , \nonumber \\
          P_{e\mu}&=&4s_{13,m}^{2}c_{13,m}^{2}s_{23}^{2}S^2_{31} \, ,\nonumber \\
          P_{e\tau}&=&4s_{13,m}^{2}c_{13,m}^{2}c_{23}^{2}S^2_{31}\, ,\nonumber \\
        P_{\mu\mu}&=&1-4s_{13,m}^{2}c_{13,m}^{2}s_{23}^{4}S^2_{31}- \nonumber \\
        &&  4s_{13,m}^{2}s_{23}^{2}c_{23}^{2}S^2_{21}-
           4c_{13,m}^{2}s_{23}^{2}c_{23}^{2}S^2_{32}\, , \nonumber \\
          P_{\mu\tau}&=&4s_{13,m}^{2}s_{23}^{2}c_{23}^{2}S^2_{21}+ \nonumber \\
        && 4c_{13,m}^{2}s_{23}^{2}c_{23}^{2}S^2_{32}- 
       4s_{13,m}^{2}c_{13,m}^{2}s_{23}^{2}c_{23}^{2}S^2_{31}\, ,\nonumber \\
        P_{\tau\tau}&=&1-4s_{13,m}^{2}c_{13,m}^{2}c_{23}^{4}S^2_{31}- \nonumber\\
    && 4s_{13,m}^{2}s_{23}^{2}c_{23}^{2}S^2_{21}-
        4c_{13,m}^{2}s_{23}^{2}c_{23}^{2}S^2_{32}\, .
\label{eq19}
\eea    
In the above,
\bea 
s_{ij,m} &=& \sin \Theta_{ij,m}\, , \nonumber \\
c_{ij,m} &=& \cos \Theta_{ij,m} 
\label{eq20}
\eea
with $\Theta_{ij,m}$ being the mixing angle in matter. Also
\bea
S_{ij}&=&\sin\l(\Delta \mu _{ij}^{2}\frac{L}{4E}\r) 
\label{eq21}
\eea
with mass square difference in matter $\Delta \mu_{ij}$ are given by 
\bea
\Delta \mu _{31}^{2}&=&\Delta m_{32}^{'2}
\frac{\sin(2\Theta_{13})}{\sin(2\Theta_{13,m})} \, ,\nonumber \\
\Delta \mu _{32}^{2}&=&\frac{\Delta m_{32}^{'2}}{2}
\l(\frac{\sin(2\Theta_{13})}{\sin(2\Theta_{13,m})}+1\r)+EV_{cc} \, ,\nonumber \\
\Delta \mu _{21}^{2}&=&\frac{\Delta m_{32}^{'2}}{2}
\l(\frac{\sin(2\Theta_{13})}{\sin(2\Theta_{13,m})}-1\r)-EV_{cc}\, .
\label{eq22}
\eea
The mixing angle $\Theta_{13,m}$ is given by 
\bea
\sin(2\Theta_{13,m})=\frac {\sin(2\Theta_{13})}
{\sqrt{(\cos(2\Theta_{13})-2EV_{cc}/\Delta m_{32}^{'2})^{2}+
\sin^2(2\Theta_{13})}}\, .
\label{eq23}
\eea
    
In this work we study the velocity induced
flavour oscillation for neutrinos. In the absence of this 
velocity induced oscillation, the oscillation length for
usual mass flavour oscillation $\sim E$, whereas
for purely velocity induced oscillation the oscillation length
$\sim 1/E$. 
Here we have an oscillation scenario where both the types
are treated in a single framework. 

\subsection{Velocity Induced Neutrino Oscillation Without MSW Effect}
\begin{figure}[h!]
\centering
\subfigure[Probability $P_{ee}$ as a function of energy at L=7000Km.]{
{\includegraphics[width=2.0in]{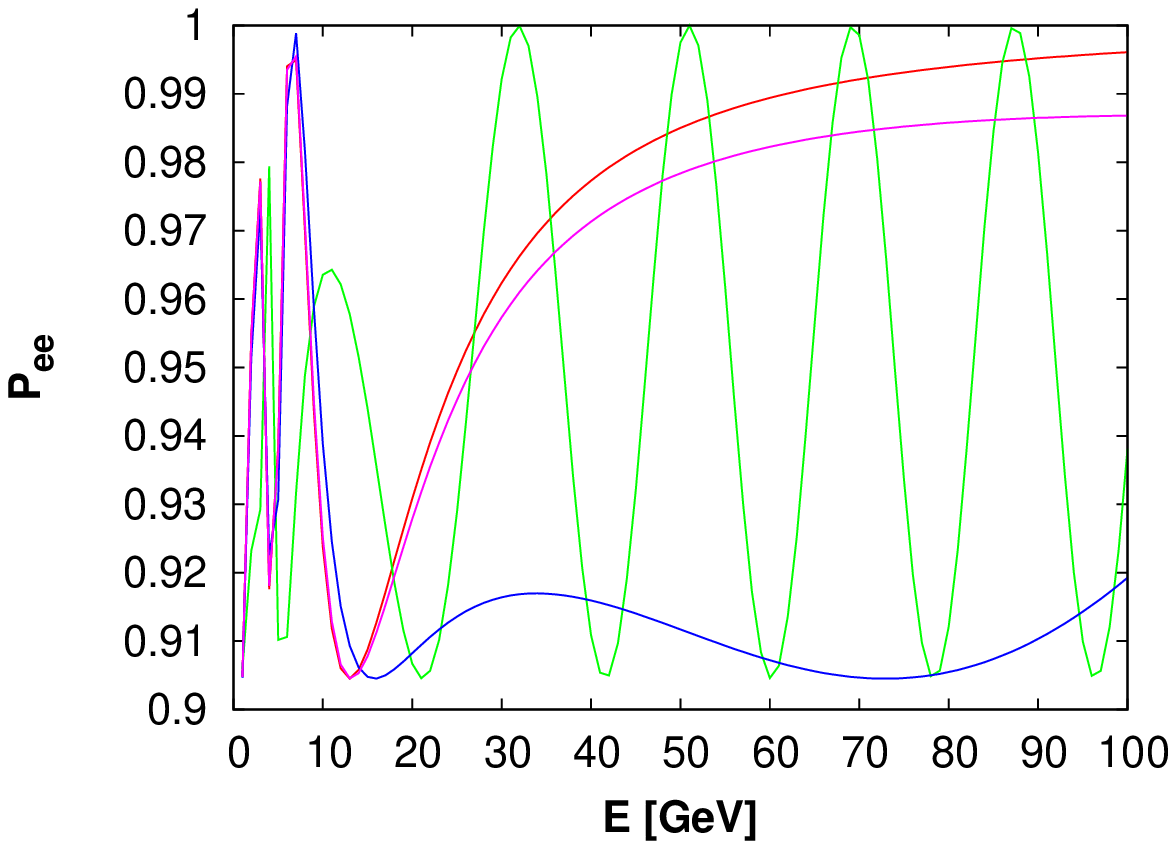}}}
\subfigure [Probability $P_{e\mu}$ as a function of energy at L=7000Km.]{
{\includegraphics[width=2.0in]{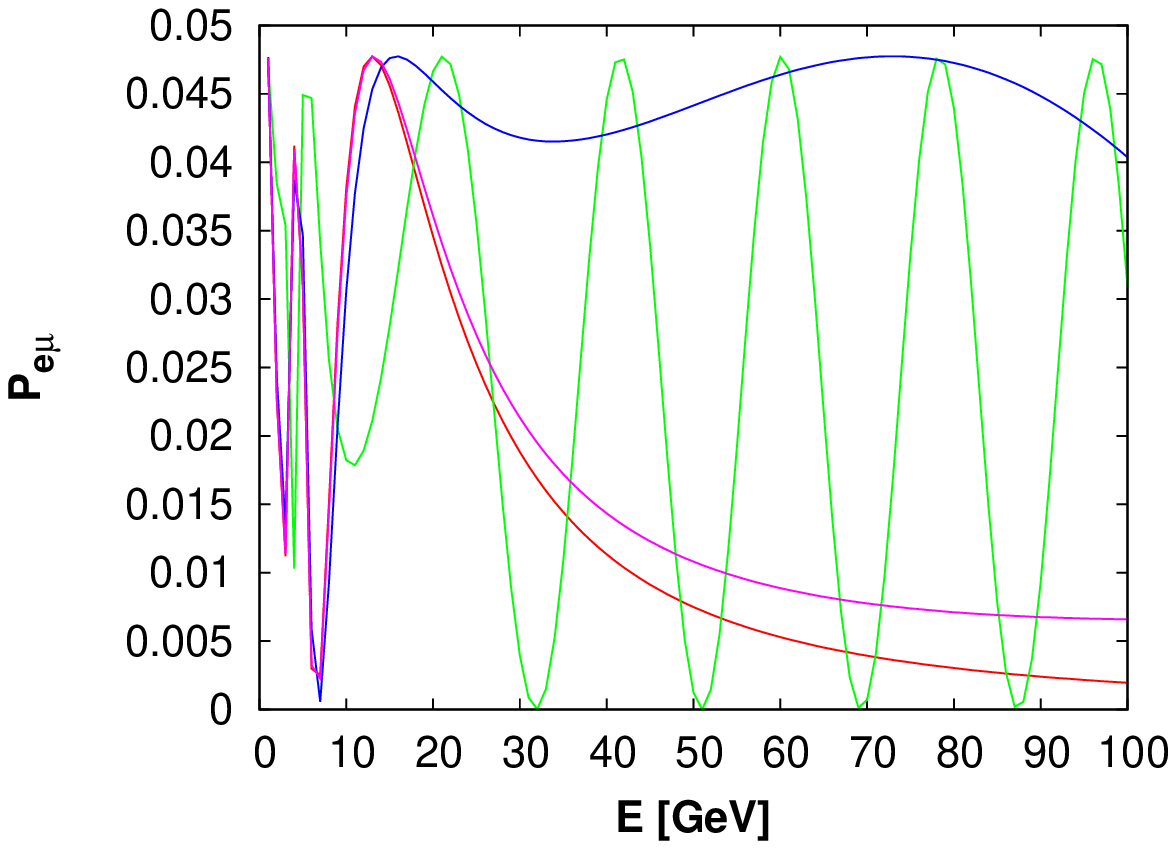}}}
\subfigure[Probability $P_{e\tau}$ as a function of energy at L=7000Km.]{
{\includegraphics[width=2.0in]{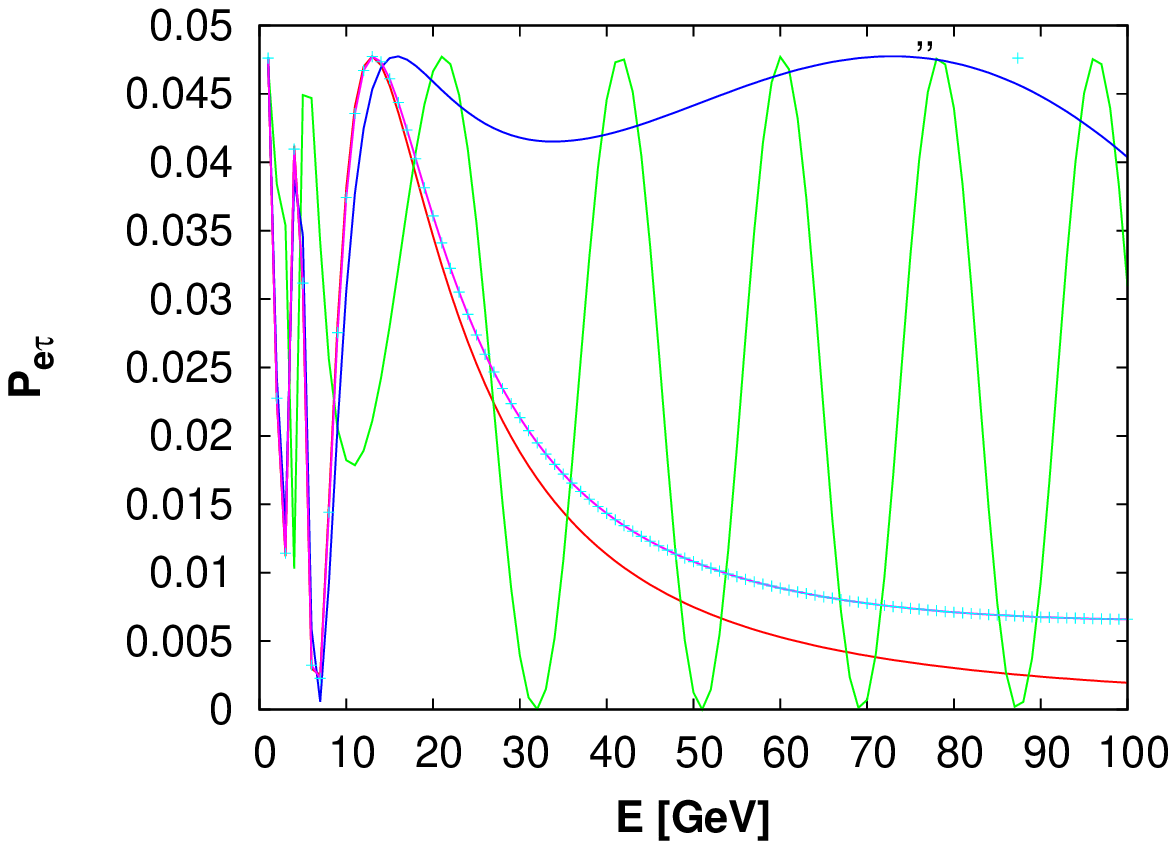}}}
\subfigure[Probability $P_{\mu \mu}$ as a function of energy at L=7000Km.]{
{\includegraphics[width=2.0in]{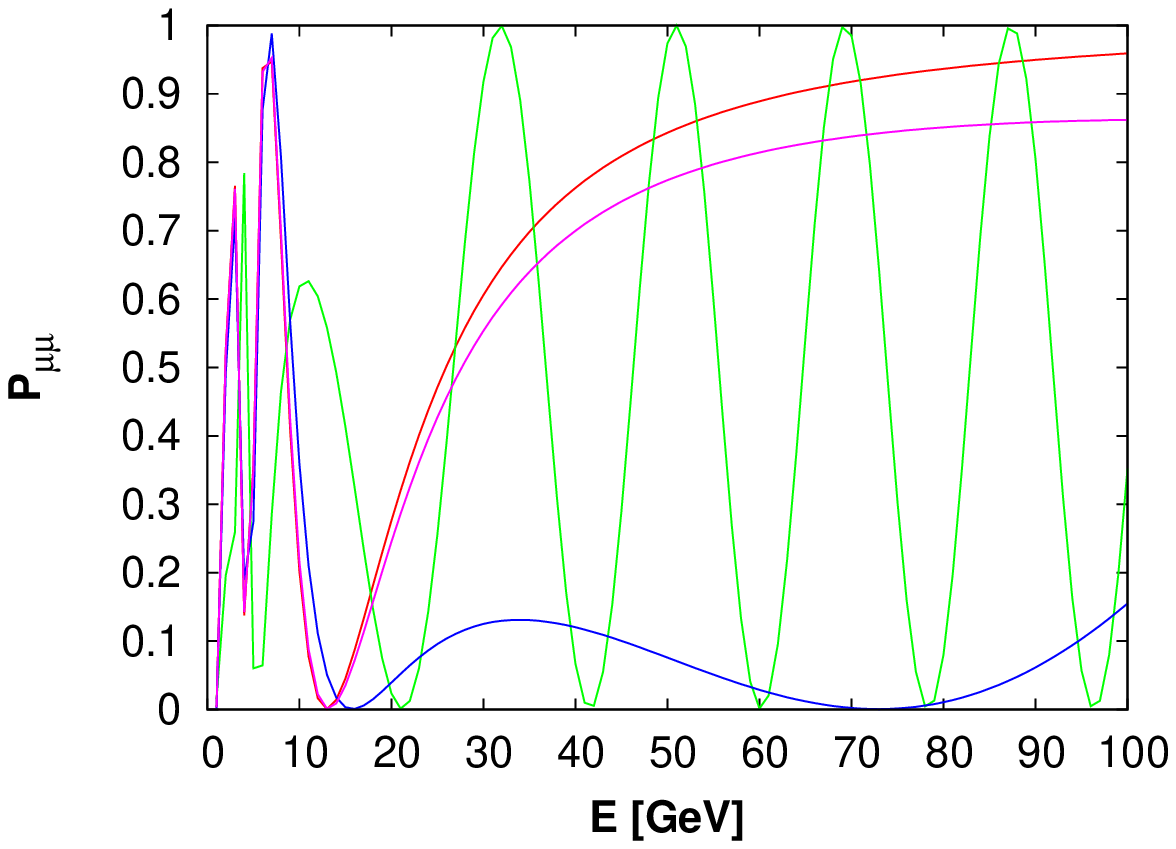}}}
\subfigure[Probability $P_{\mu \tau}$ as a function of energy at L=7000Km.]{
{\includegraphics[width=2.0in]{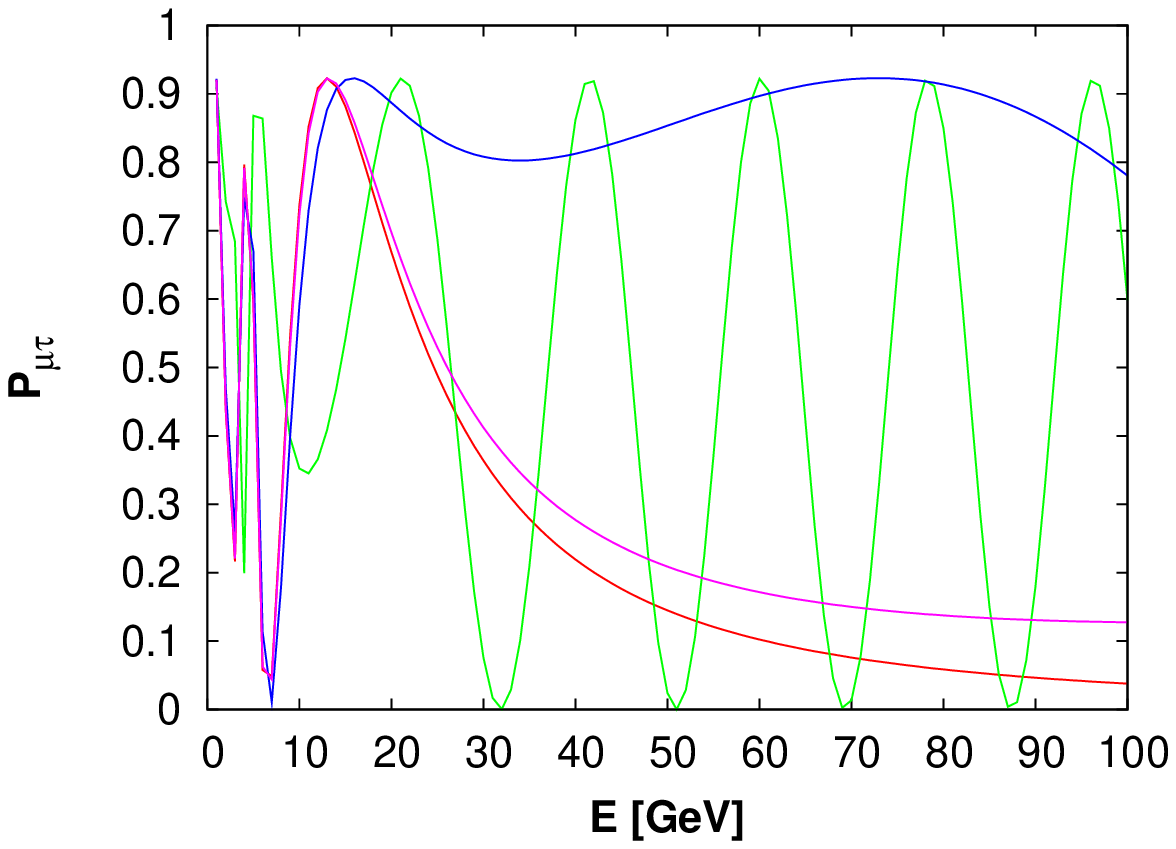}}}
\subfigure[Probability $P_{\tau \tau}$ as a function of energy at L=7000Km.]{
{\includegraphics[width=2.0in]{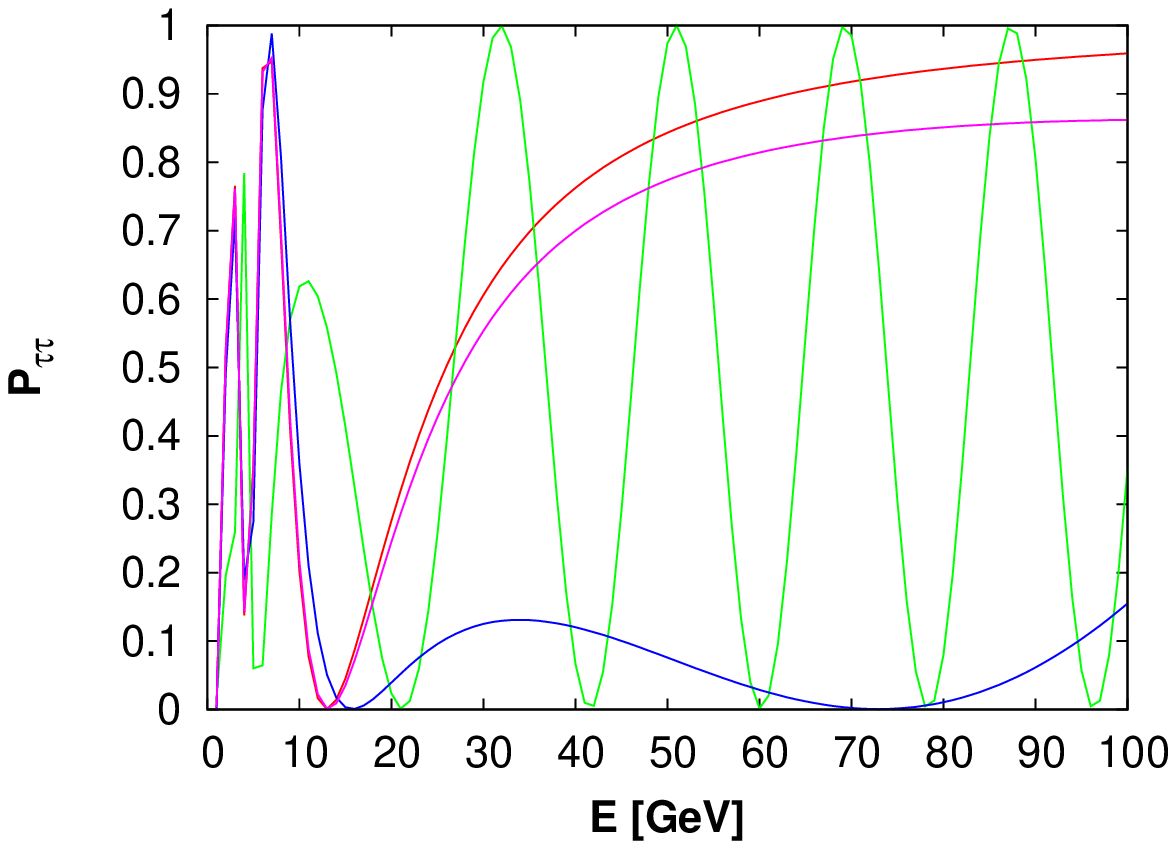}}}
\caption{Velocity induced neutrino oscillation probabilities in vacuum
(without MSW effect) and 
their comparisons when the effect of velocity induced oscillations are
not present. The red lines in all the plots (a - f) are for the cases when
velocity effect on the probabilities are not considered. The green, blue
and pink lines represent the probabilities with velocity effects for
$\Delta V = 10^{-23},\, 10^{-24},\, 10^{-25}$ respectively in all the plots
(a-f).  }
\label{fig1}
\end{figure}

In the oscillation framework discussed above, we will first consider
the velocity mediated oscillation scenario in absence of matter.
This means, in this section, we explore the oscillation probabilities
for the case when both the mass eigenstates and velocity eigenstates
are different from flavour eigenstates of the neutrinos without 
including MSW effect.
This can be realised
by setting the terms for matter effect to zero in Eqs.~(\ref{eq12}) 
- (\ref{eq23}). The oscillation 
parameters, $\Delta m_{ij}^2$ and vacuum mixing angles $\Theta_{ij}$ 
($i,j \equiv 1,2,3$) for vacuum oscillation 
(in Eqs.~(\ref{eq18} - \ref{eq22}))
are taken to be  $\Delta m_{32}^{2} = 2.3 \times 10^{-3}$ eV${^2}$  
and mixing angles are $\Theta_{12} = 33.21^{\rm o}$, 
$\Theta_{23} = 45^{\rm o}$ and
$\Theta_{13} = 9^{\rm o}$ respectively \cite{fogli1,thakore}.
 The probabilities are calculated 
for a reference baseline length of 7000 Km with above mentioned
values of oscillation parameters and for three different demonstrative 
values of the parameter $\Delta V \equiv \Delta V_{23}$
(the difference in neutrino velocities)
namely  $\Delta V = 10^{-23}$,
$\Delta V = 10^{-24}$ and $\Delta V = 10^{-25}$
and then they are compared 
with the case of
ordinary vacuum oscillation for massive neurinos in which 
velocity mediated oscillations are neglected 
($\Delta m^{'2}_{32} = \Delta m^{2}_{32}$ etc.). 
The bounds on $\Delta V$ from previous studies are discussed 
in Sect. 1 (Introduction).
The results are 
shown in the six plots of Fig.~\ref{fig1} in the sequence,
Fig.~\ref{fig1}a - Fig.~\ref{fig1}f,
for the variations of different
probabilities $P_{ee}$, $P_{e\mu}$, $P_{e\tau}$,  $P_{\mu\mu}$,
$P_{\mu\tau}$,  $P_{\tau\tau}$ respectively
with neutrino energies $E$.
Here $P_{\alpha_1 \alpha_2}$ denotes the probability of oscillation 
of a neutrino $|\nu_{\alpha_1}\rangle$
of flavour $\alpha_1$ to a neutrino $|\nu_{\alpha_2}\rangle$ 
of flavour $\alpha_2$ and the symbols $\alpha_1,\alpha_2$ signify
$\alpha_1,\alpha_2 \equiv e,\mu,\tau$ as the case may be.
In each of the plots of Fig.~\ref{fig1}, the red line represents
the vacuum oscillation probability without velocity mediated oscillation 
whereas the green, blue and pink lines represent  
both mass and velocity induced flavour oscillation probabilities for three 
different values of $\Delta V$ namely $\Delta V = 10^{-23}$, 
$\Delta V = 10^{-24}$ and $\Delta V = 10^{-25}$ respectively. It is clear
that the neutrino flavour oscillations induced by the difference in 
velocities of two neutrinos modify the normal flavour oscillations 
for massive neutrinos (only mass induced flavour oscillation),
for all the cases considered. In Fig.~\ref {fig1}a, the survival 
probability of an electron neutrino ($\nu_e$) is plotted for
various neutrino energies $E$. 
It is seen from Fig. \ref{fig1}a that for $E \la 13$ GeV, the 
probabilities with $\Delta V = 10^{-25}$ and $\Delta V = 10^{-24}$
are comparable with the  case for which the velocity effect 
is not considered (red line).
Beyond $E \sim 13$ GeV, while the case with
$\Delta V = 10^{-25}$ is comparable with that without velocity effect (red
line), the 
probability deviates significantly for $\Delta V = 10^{-24}$. The case for 
$\Delta V = 10^{-23}$ however, 
is distinctly different from all the cases in this scenario for the whole 
range of energy considered here. All the probability 
plots in Fig. \ref{fig1}a-\ref{fig1}f
exhibit these trends. Fig.~\ref{fig1}a - \ref{fig1}f 
clearly demonstrates that for $\Delta V = 10^{-23}$ and 
$\Delta V = 10^{-24}$, the effect of velocity-flavour mixing is 
more pronounced and for the former case (higher $\Delta V$) one can  
observe multiple the oscillatory behaviour (the green lines). Thus, 
in case the neutrinos of different species indeed differ in velocities
such that $\Delta V$ is of the relevant order, one should observe the 
effect of velocity mediated oscillations for the neutrinos 
traversing a baseine length which in the present cases taken to be 
$\sim 7000$ Km for demonstration.    

\subsection{Velocity Induced Neutrino Oscillation in Matter: MSW 
Effect and Resonance}

\begin{figure}[h!]
\centering
\subfigure[Probability $P^{m}_{ee}$ as a function of energy at L=7000Km.]{
\includegraphics[width=2.0in]{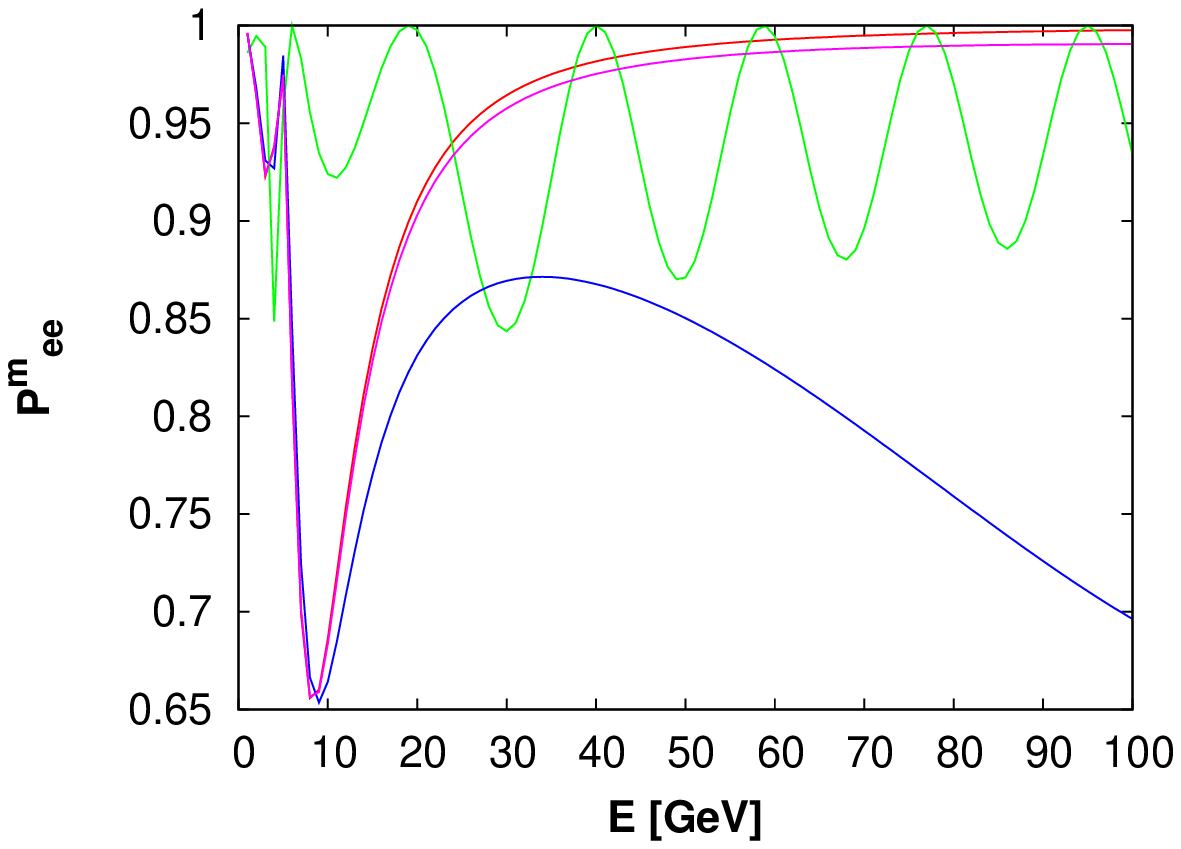}}
\subfigure [Probability $P^{m}_{e\mu}$ as a function of energy at L=7000Km.]{
\includegraphics[width=2.0in]{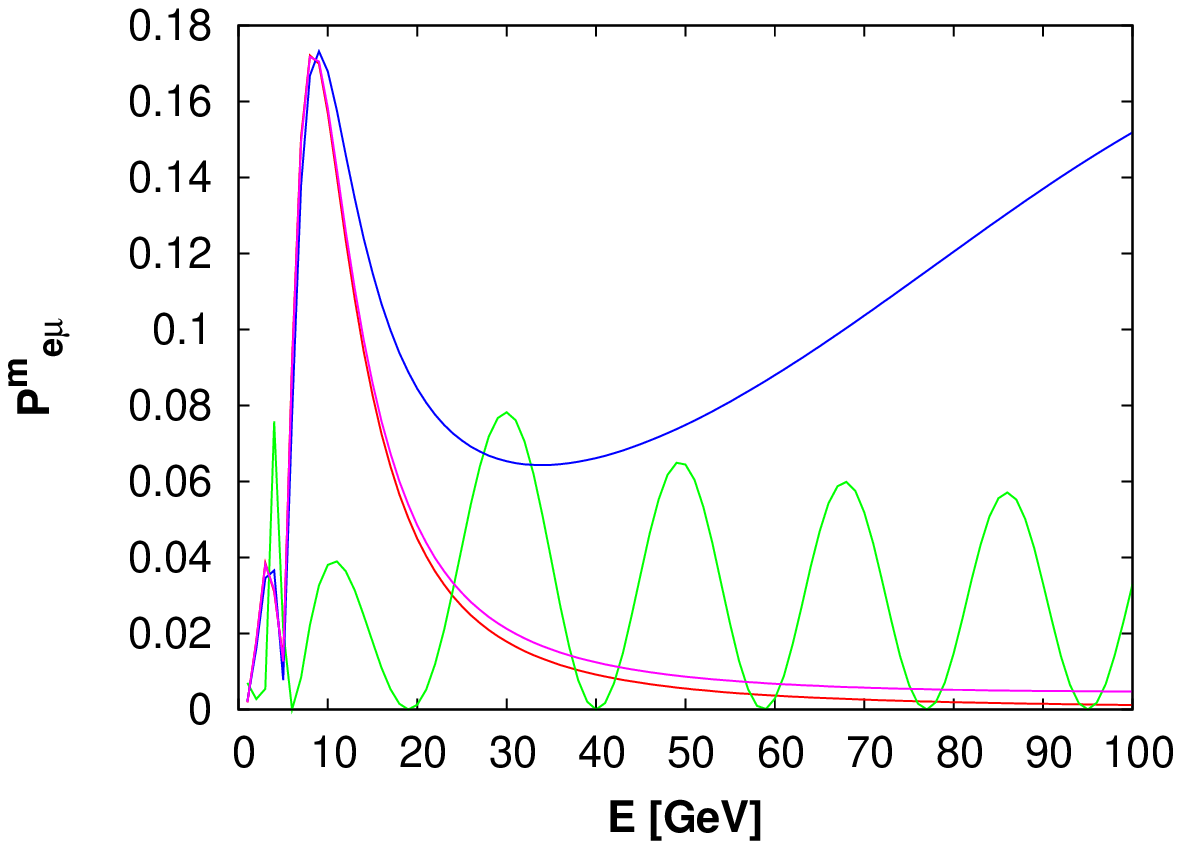}}
\subfigure[Probability $P^{m}_{e\tau}$ as a function of energy at L=7000Km.]{
\includegraphics[width=2.0in]{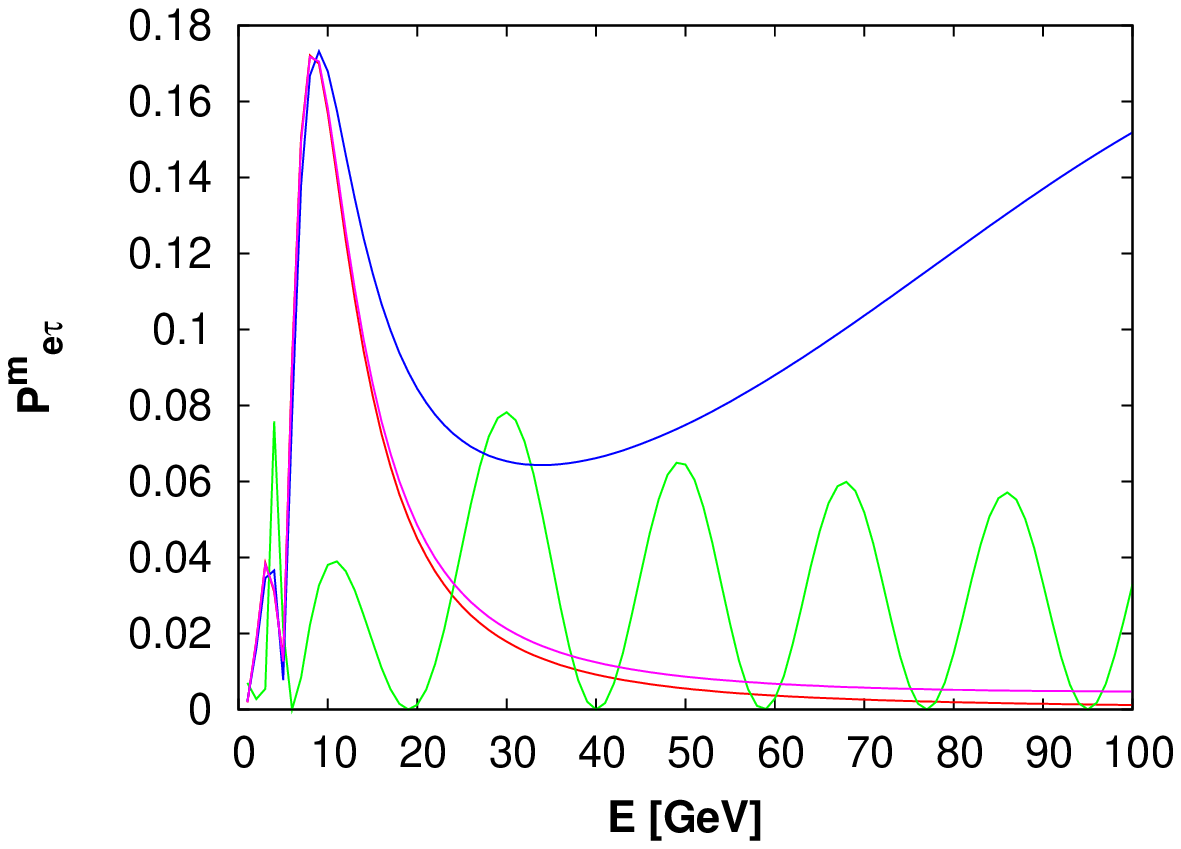}}
\subfigure[Probability $P^{m}_{\mu \mu}$ as a function of energy at L=7000Km.]{
\includegraphics[width=2.0in]{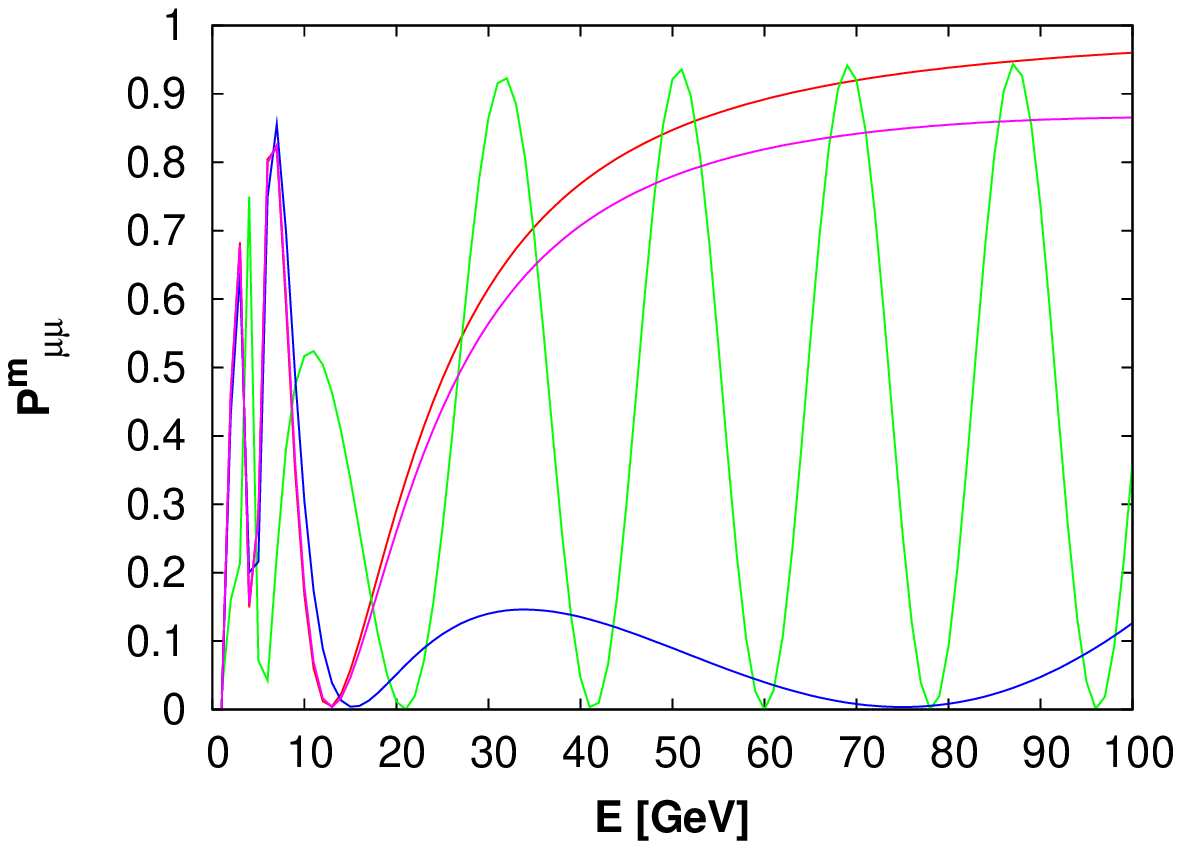}}
\subfigure[Probability $P^{m}_{\mu \tau}$ as a function of energy at L=7000Km.]{
\includegraphics[width=2.0in]{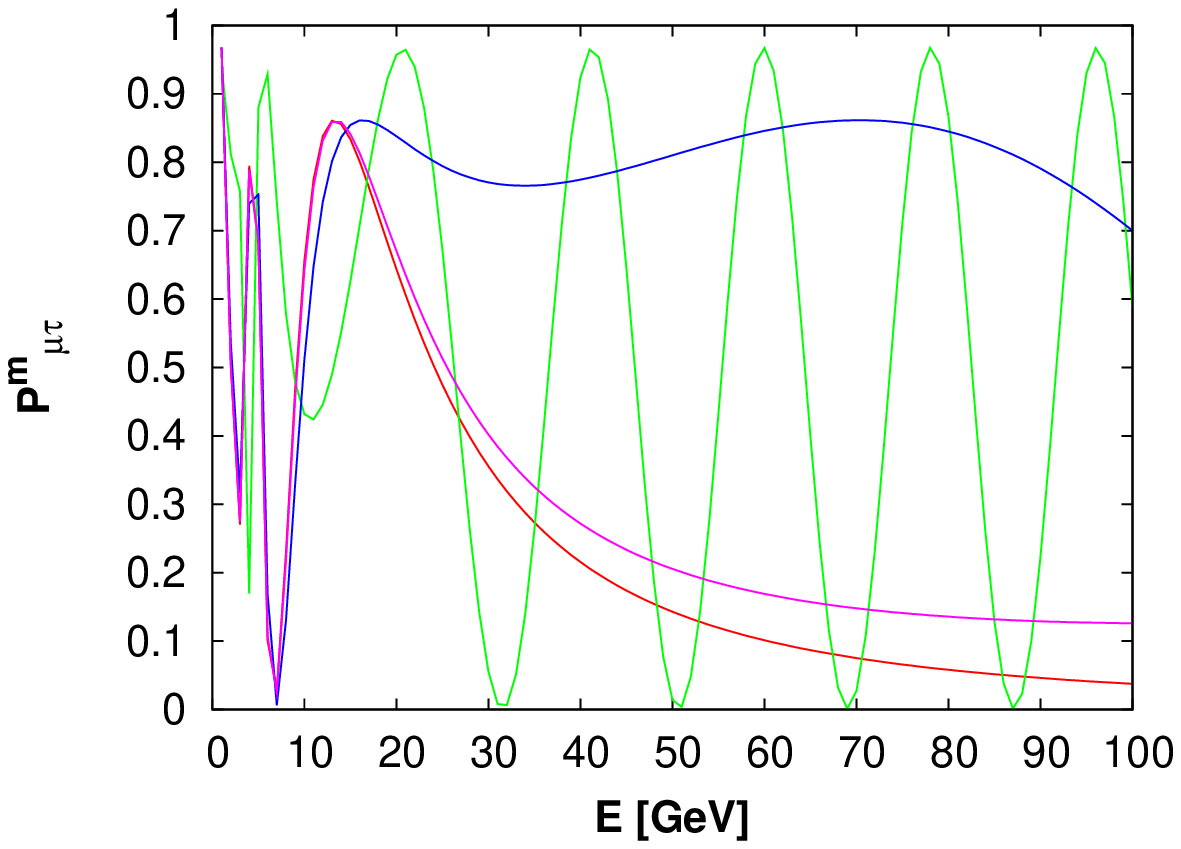}}
\subfigure[Probability $P^{m}_{\tau \tau}$ as a function of energy at L=7000Km.]{
\includegraphics[width=2.0in]{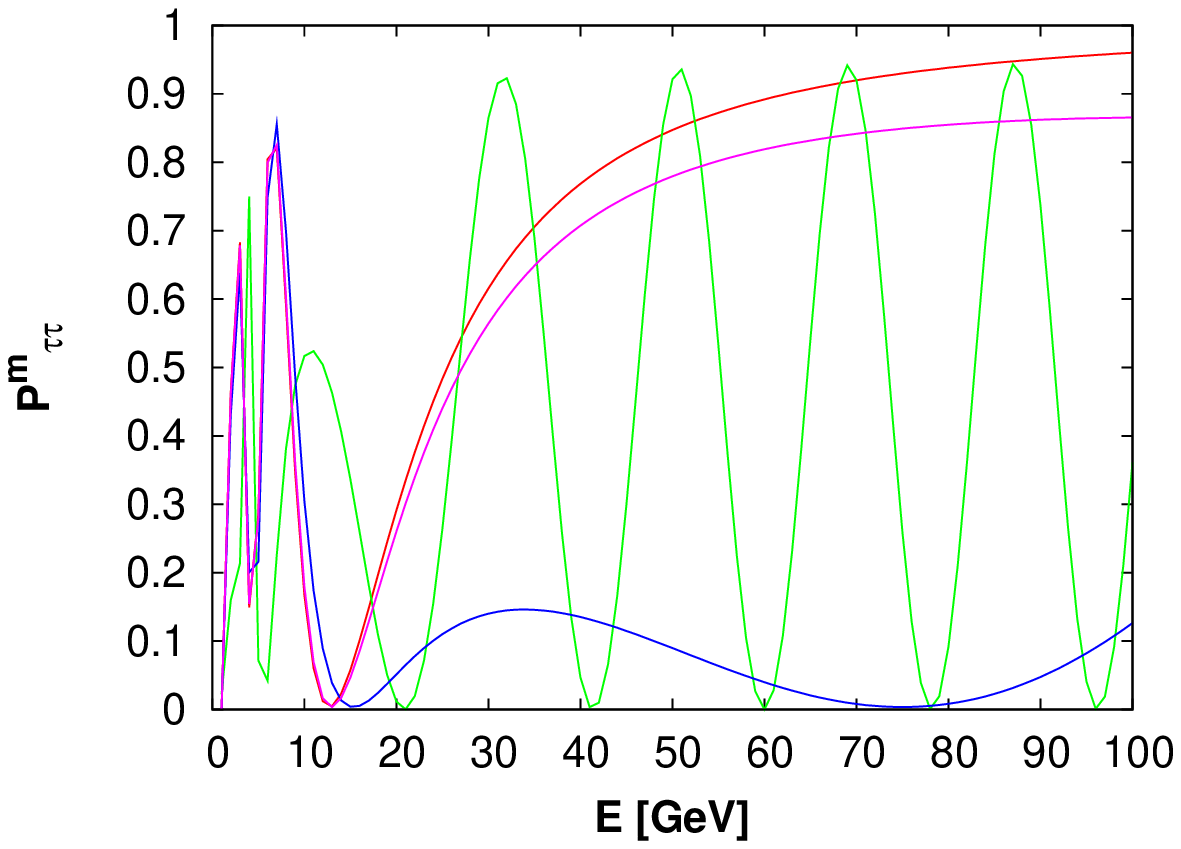}}
\caption{Velocity induced neutrino oscillation probabilities in matter and 
their comparisons when the effect of velocity induced oscillations are 
not present. The red lines in all the plots (a - f) are for the cases when 
velocity effect on the probabilities are not considered. The green, blue 
and pink lines represent the probabilities with velocity effects for 
$\Delta V = 10^{-23},\, 10^{-24},\, 10^{-25}$ respectively in all the plots
(a-f).  }
\label{fig2}
\end{figure}

The passage of neutrinos through matter can considerably modify the 
neutrino oscillation probabilities. We study in this section 
how the velocity induced vacuum oscillation probabilities
discussed earlier are modified when the matter effect (or MSW effect)
is included. In the present context, this may be useful 
if one considers the case of baseline neutrinos from an accelerator 
or a neutrino factory propagating from their origin to a detector 
through the earth matter. 
In order to demonstrate the effect of different $\Delta V$
on the neutrino oscillation probabilities with matter effect, we
have chosen a mean earth matter density of 4.15 gm/cc while 
the neutrinos traverse a representative baseline length of 
7000 Km through the earth matter.
The probabilities at different neutrino energies $E$ 
are calculated using Eqs.~(\ref{eq17}) - (\ref{eq23})  
and are plotted in Fig.~\ref{fig2}a - \ref{fig2}f.
As in Fig.~\ref{fig1}, here too $P_{\alpha_1 \alpha_2}$ represents 
the oscillation
probability of a neutrino from flavour $\alpha_1$ to a flavour $\alpha_2$.

Fig. 2 demonstrates how the velocity induced oscillations affect
the usual mass-flavour neutrino oscillation probabilities in presence
of matter. For example, the electron neutrino survival probabilities
$P^{m}_{ee}$ in Fig. \ref{fig2}a, although appears similar 
for the cases when $\Delta V =0$ and $\Delta V = 10^{-25}$,
the effect of 
$\Delta V$ shows up when $\Delta V = 10^{-24}$ beyond  
$E \sim 15$ GeV. On the other hand, for $\Delta V = 10^{-23}$, 
the variations of 
$P^{m}_{ee}$ with $E$  differ significantly not only 
from the case when velocity effect is absent ($\Delta V = 0$) 
but also from the other two probabilities for two other 
$\Delta V$ values as well. 
This may also be noted that the shapes of $P^{m}_{ee}$ 
with matter effect (Fig.~\ref{fig2}) differ from those
without matter effect ($P_{ee}$ in Fig.~\ref{fig1}) and these 
differences are more evident for the cases when $\Delta V = 10^{-23}$ and 
$\Delta V = 10^{-24}$. The variations of all the probabilities 
shown in Figs. \ref{fig2}a - \ref{fig2}f exhibit similar trends. Similar to 
vacuum oscillation scenario (without matter effect), here too the 
variations of probabilities
exhibit multiple oscillatory behaviour for 
$\Delta V = 10^{-23}$ although the natures of variations are notably 
different.   

From the plots (a - f) in Fig. \ref{fig2}, this can be arguably stated that 
the effects of $\Delta V$ on the oscillation probabilities are indeed 
significant and this is most evident for $\Delta V = 10^{-24}$ and 
$\Delta V = 10^{-23}$ among the cases considered here. 
As mentioned earlier, for the latter case,  the probabilities at 
different neutrino energies show a distinct multiple
oscillatory nature in contrast to other probabilities. If many 
oscillations are accommodated in the oscillation length for the 
considered scenario then the probability would be averaged out.     
Comparison of Fig.~\ref{fig2} with Fig.~\ref{fig1} also clearly demonstrates
that the matter effect modifies
the vacuum probabilities (without matter effects).  
As an example, for $\Delta V = 10^{-25}$ the value of $P^{m}_{e\tau}$, 
at a reference neutrino energy $E = 10$ GeV, 
changes from 0.04 (Fig.~\ref{fig1}) for no matter effect to 0.16 for the
cases with matter effect (Fig.~\ref{fig2}). 

\begin{figure}[h!]
\centering
\subfigure[ $ \cos(2\Theta_{13}) - 2EV_{cc}/\Delta m_{32}^{'2}$ vs E for $\rho = 4.00 $ .]{
\includegraphics[width=2.0in]{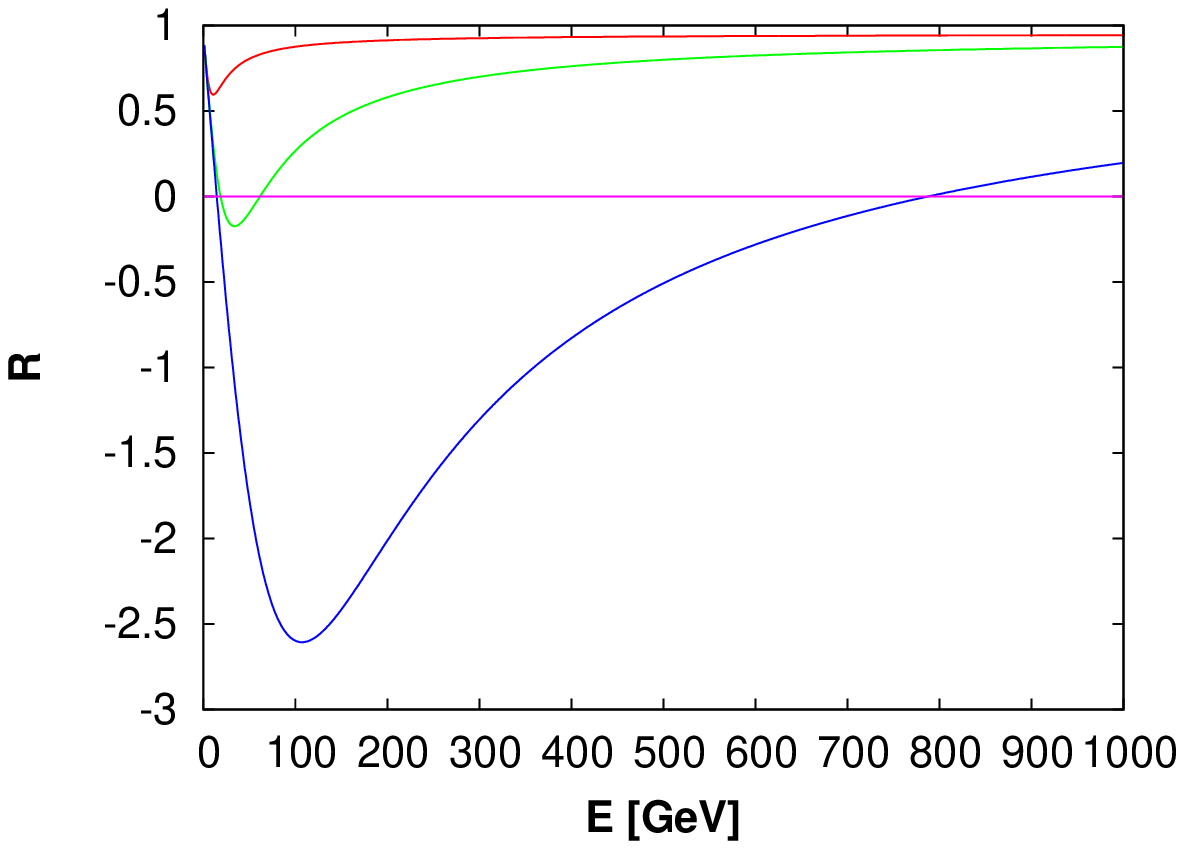}}
\subfigure[ $ \cos(2\Theta_{13}) - 2EV_{cc}/\Delta m_{32}^{'2}$ vs E for $\rho = 4.15$ .]{
\includegraphics[width=2.0in]{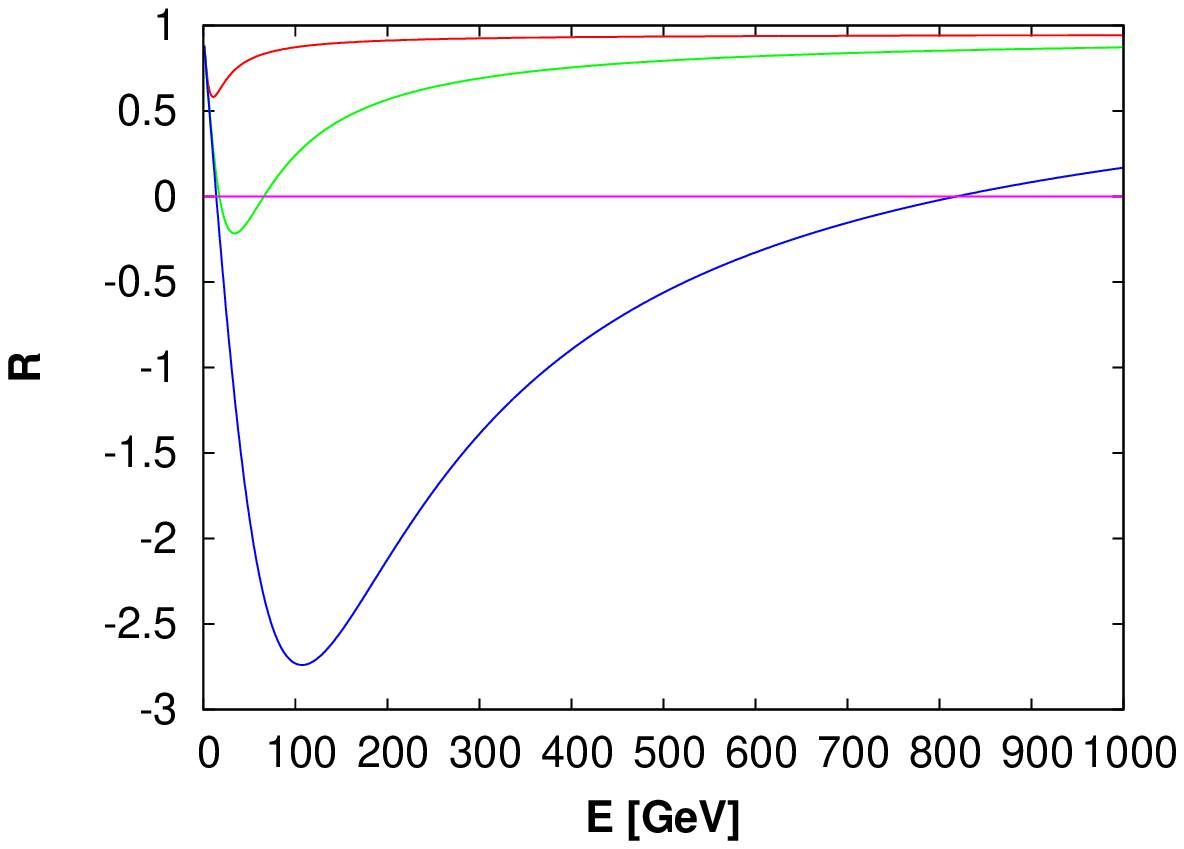}}\\
\subfigure[ $ \cos(2\Theta_{13}) - 2EV_{cc}/\Delta m_{32}^{'2}$ vs E for $\rho = 4.50 $ .]{
\includegraphics[width=2.0in]{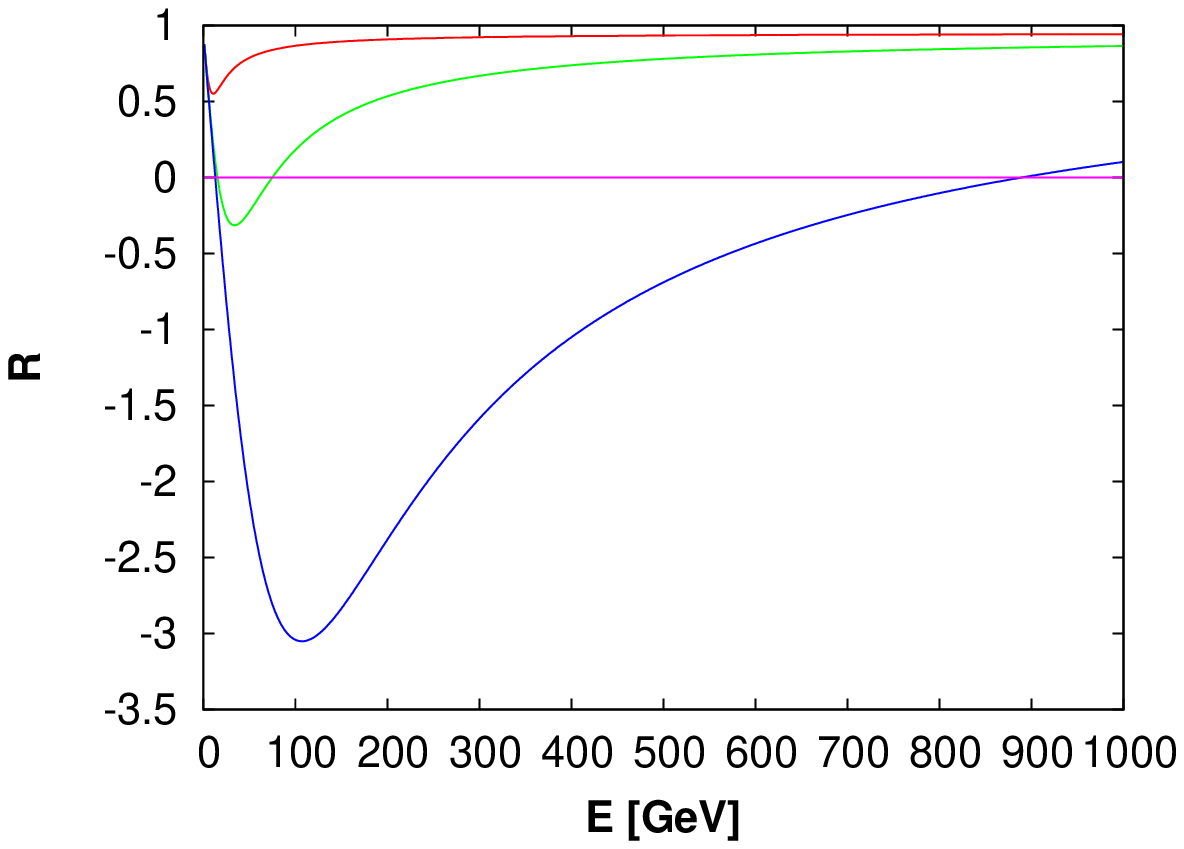}}
\caption{Effect of matter density $\rho$ on MSW resonance for different values of $\Delta V$}
\label{fig3}
\end{figure}

We also investigate the phenomenon of resonance which gives 
the maximal mixing between two neutrinos under the oscillation framework
with matter effect.
The possible occurence of resonance can be understood from 
Eq.~(\ref{eq23}) where the mixing angle $\Theta_{13,m}$ will be maximum
when $R = \cos(2\Theta_{13}) - 2EV_{cc}/\Delta m_{32}^{'2} = 0$.
Therefore, from Eq.~(\ref{eq23}), $\Theta_{13,m} = \frac{\pi}{4}$  
at resonance. With $V_{cc} = \sqrt {2} G_F N_e$ (Eq.~\ref{eq13}),
the resonance condition is 
\bea
\sqrt {2} G_F N_e = \frac {\Delta m_{32}^{'2}} {2E} \cos(2\Theta_{13})\,.
\label{eq24}
\eea 
The electron density $N_e$ in Eq.~(\ref{eq24}) can be obtained as
$N_e = \rho N_A Y_e$, where $\rho$ is the density (the density of the 
earth in our case of long baseline neutrino), $N_A$ is the Avogadro number 
and $Y_e$ is the electron fraction.  
We calculate the variation of the quantity $R$ for three chosen 
values of matter densities namely $\rho =4.0,~4.15,~4.5$ gm/cc
and for the three values of $\Delta V$ considered in this work. 
The results are plotted in Fig. \ref{fig3}(a - c). For all the 
calculations we restrict ourselves to the case of normal mass 
hierarchy for the neutrinos whereby $\Delta m_{32}^{'2} > 0$ 
($\Delta m_{32}^2 > 0$ as well).  

A discussion is in order. Using the expression for $\Delta m_{32}^{'2}$
from Eq.~(\ref{eq18}), and substituing $\Delta V$ for $\Delta V_{32}$ 
as is done in this work, the expression for $R$ is given as
\bea
R &=& \cos(2\Theta_{13}) - \frac {2\sqrt {2} G_F N_e E} {\Delta m^2_{32}
+ 2\Delta V E^2}\, . 
\label{eq25a}  
\eea
Fig.~\ref{fig3}a - \ref{fig3}c 
shows that for any given non-zero value of $\Delta V$, the $R$ vs $E$ 
plots originate from a finite value 
($\cos(2\Theta_{13})$) and then after reaching a minimum, asymptotically 
approach to the $\cos(2\Theta_{13})$ value for larger values of $E$. 
This behaviour can be easily understood from Eq.~(\ref{eq25a}). 
The $E$ dependence of $R$ comes from the second term 
$\frac {2\sqrt {2} G_F N_e E} {\Delta m^2_{32}
+ 2\Delta V E^2} ( = X,~{\rm say})$
of the RHS 
of Eq.~(\ref{eq25a}) through the combined effect of $E^2$ at the  
denominator (quadratic inverse) and $E$ (linear) at the numerator.
Thus for low values of $E$, the quantity $R$ will decrease because 
of the negative term $X$ on the RHS of Eq.~(\ref{eq25a}).
But as $E$ increases, the $E^2$ in $X$ starts dominating over the 
linear $E$ dependence of $X$ and since $E^2$ is at the denominator, 
$X$ decreases faster than 
its increase with $E$, due to the linear $E$ dependence. This clearly explains 
the nature of $R$ vs $E$ plots in Fig.~\ref{fig3} discussed above. 
We mention in the passing that from Eq.~(\ref{eq25a}), it is also evident that
$R$ vs $E$ plot will be a straight line when $\Delta V = 0$ (no velocity 
effects).  
\par
The resonance condition of Eq.~(\ref{eq24}) takes the 
form 
\bea
\sqrt {2} G_F N_e &=& \left ( \frac {\Delta m_{32}^{2}} {2E} 
+ \Delta V E \right ) \cos(2\Theta_{13}) \nonumber \\
&=& \left ( \frac {\Delta m_{32}^{2} + 2 \Delta V E^2} {2E} \right ) 
\cos(2\Theta_{13})\, .
\label{eq25}
\eea
Eq.~(\ref{eq25}) can be cast in the form 
\bea
AE^2 - BE + C & = & 0 \nonumber \\
A = 2 \Delta V \cos(2\Theta_{13}),~~  
B = 2 \sqrt{2} G_F N_e\,, &&
C = \Delta m_{32}^{2} \cos(2\Theta_{13}).
\label{eq26}
\eea
The solutions of $E$ in Eq.~(\ref{eq26}) give the resonance energy $E_R$
for which the resonance condition (Eq.~(\ref{eq24}) or $R=0$) is satisfied. 

The two solutions of $E$ in Eq.~(\ref{eq26}) are given by
\bea
E &=& \frac {B \pm \sqrt {B^2 - 4AC}} {2A}\, .
\label{eq27}
\eea 
Eq.~(\ref{eq27}) has two real solutions for E when 
$B \geq \sqrt {B^2 - 4AC}$ and $B^2 > 4AC$. Thus we obtain two 
resonance energies $E_R$ when these conditions are satisfied. 
On the other hand, $E$ will have only one solution when 
the condition $B^2 =4AC$ is satisfied. In that case the only solution 
will be $E = E_R = \frac{B} {2A}$ and imposing the condition 
$B^2 = 4AC$, one readily obtains, as the single soluion, 
$E_R = \sqrt {\frac {C} {A}}$.  
For 
computation of the quantities $A$, $B$ and $C$ (Eq.~(\ref{eq26})) 
it is to be noted that 
$B$ is a constant for a chosen matter of known matter density $\rho$ 
through which the neutrinos traverse, $C$ is also 
a constant since the oscillation parameters $\Delta m_{32}^{2}$ and 
$\Theta_{13}$ are obtained from the neutrino experiments.  
For the computation 
of $A$, while the factor $\cos(2\Theta_{13})$ is fixed by the experimental
value of $\Theta_{13}$, 
only the unknown velocity-induced oscillation parameter $\Delta V$  
is varied (in the present work, as mentioned, three values are chosen 
for the same). This is also true for other conditions for two real solutions
of the resonance energy (as also for obtaining no solutions). 
Therefore, the parameter $\Delta V$ dictates 
the various solutions of resonance energy for a given matter density.   

One observes from the Fig. \ref{fig3}a - \ref{fig3}c that 
for $\Delta V = 10^{-23}$ no resonance occurs. This means that for 
the chosen values of the matter densities, the resonance condition 
(Eq.~\ref{eq25}) is not satisfied at all.
The quantity $R$, for this $\Delta V$ value however passes through a minima
and then approaches to its initial value of $\cos(\Theta_{13})$ as discussed
above. 
For both the cases, when $\Delta V = 10^{-24}$ and $\Delta V = 10^{-25}$, 
the resonances 
occur twice at two different energy values. The difference between 
the two resonance energies should increase with the decrease of 
$\Delta V$ values and this also can be explained from the difference
of two solutions in Eq.~(\ref{eq27}) which is 
$E_{r_2} - E_{r_1} = \sqrt { \frac {B^2} {A^2} -4 \frac{C} {A} }$.  
Thus the effect of velocities-induced oscillations of neutrinos can also be  
manifested in the occruence of two resonances at two different energies. 
The results are summerised in Table 1.
\begin{table}[h]
\tbl{{ Resonance energy $E_r$ for three different matter 
density $\rho$ and three 
diffrent values of $ \Delta V $ }}
{\begin{tabular}{@{}ccc@{}} \toprule
$ \rho~(N_{A}/cm^{3}) $  &    $\Delta V$ & Resonance energy $E_r$ in GeV \\
\colrule
         \hphantom{00}& \hphantom{0}1E-23 & \hphantom{0}  No resonance    \\
  4.00   \hphantom{00}& \hphantom{0}1E-24 & \hphantom{0}18.70,\hphantom{0}61.50        \\      
         \hphantom{00}& \hphantom{0}1E-25 & \hphantom{0}14.60,\hphantom{0}787.70      \\
\hline
         \hphantom{00}& \hphantom{0}1E-23 & \hphantom{0}  No resonance      \\
  4.15   \hphantom{00}& \hphantom{0}1E-24 & \hphantom{0}17.50,\hphantom{0}65.70        \\      
         \hphantom{00}& \hphantom{0}1E-25 & \hphantom{0}14.10,\hphantom{0}818.30      \\
\hline
         \hphantom{00}& \hphantom{0}1E-23 & \hphantom{0}  No resonance     \\
  4.50   \hphantom{00}& \hphantom{0}1E-24 & \hphantom{0}15.40,\hphantom{0}74.90        \\      
         \hphantom{00}& \hphantom{0}1E-25 & \hphantom{0}12.90,\hphantom{0}889.60      \\   \botrule
\end{tabular}\label{ta1} }
\end{table}

However for antineutrinos, the 
interacting potenial is $-V_{cc}$. 
It is evident therefore that for antineutrinos, the resonance like 
phenomena will not occur unless $\Delta m_{32}^{2}$ is negative.

From the study of both Fig.~\ref{fig2} and Fig.~\ref{fig3}, it can be said
that among the three values of $\Delta V$ chosen for the present analysis, 
$\Delta V = 10^{-24}$ is most effective.  
The choice $\Delta V = 10^{-23}$ do not show any resonance phenomenon 
and the probability shows multiple oscillations. For the choice, 
$\Delta V = 10^{-25}$, the resonance energies are wide apart 
(Fig.~\ref{fig3}) and the probability almost coincide with that 
when velocity effects are absent.

\section{Effect of Velocity Induced Oscillation on a Long Baseline 
Neutrino Experiment}
 
We also investigate how the velocity induced oscillation affects the
neutrino yield at a detector for long baseline neutrinos.
In a long baseline neutrino experiment, the neutrinos are usually
produced in a neutrino factory where the protons are bombarded on a
target to form pions. The pions thus decay to yield muons. The muons 
are collected in a muon storage ring where they decay to produce 
muon and electron neutrinos following the 
processes $\mu^{-} \rightarrow e^{-} + \bar{\nu}_e +\nu_{\mu} $
and $\mu^{+} \rightarrow e^{+} + \nu_e +\bar{\nu}_{\mu}$. Neutrinos thus
produced in a muon storage ring of a neutrino factory are then directed to a  
neutrino detector several kilometers (long baseline) away. These neutrinos have to traverse a
distance through earth matter before reaching the detector.

The $\nu_{\mu}$($\bar \nu_{\mu}$) flux from such a neutrino factory is given by \cite{geer,donini}

\bea \frac{ d^2 \Phi_{\nu_{\mu},
\bar\nu_{\mu}} }{ dy dA}  =  \frac{ 4 n_\mu }{ \pi L^2 m_\mu^6 }
\,\,  E_\mu^4 y^2 \, (1 - \beta) \,\,  \left
[ 3 m_\mu^2 - 4  E_\mu^2 y \, (1 - \beta) \right ]  
\label{eq29}
\eea
and the $\nu_e$($\bar {\nu}_e$) flux is given by
\bea \frac{ d^2 \Phi_{\nu_e,
\bar\nu_e} }{ dy dA}  =  \frac{ 2 n_\mu }{ \pi L^2 m_\mu^6 }
\,\,  E_\mu^4 y^2 \, (1 - \beta) \,\,  \left
[  m_\mu^2 - 2  E_\mu^2 y \, (1 - \beta) \right ].  
\label{eq30}
\eea 
In Eqs. \ref{eq29} - \ref{eq30} the angle $\phi$ between the 
neutrino beam direction to the detector
and the beam axis is taken to be zero and neutrinos are assumed 
to have no polarisation. In the above the quantity $E_{\mu}$ is the 
energy of the muon, $n_{\mu}$ denotes the number of muons injected in
the storage ring, $L$ is the distance between the storage ring and 
detector, $m_{\mu}$ is the mass of the muon, $\beta$ 
is the boost factor and $y = \frac{E_{\nu}}{E_{\mu}}$, $E_{\nu}$ being 
the energy of the neutrino.
Such long baseline neutrinos from a neutrino factory will suffer matter 
induced oscillation during 
its passage through earth matter to a far away detector. 
For the present calculation, we consider the far away detector 
to be an iron calorimeter (ICAL) made up of a stack of different 
layers of iron plates.
A $\nu_{\mu}$ ($\bar{\nu}_{\mu}$), after reaching this ICAL detector 
will undergo charged current interaction with the iron of ICAL and produce
track of $\mu^-$ ($\mu^+$) as they pass through diferent layers of ICAL 
detector. If the detector is magnetised, one can distinguish
the tracks generated by $\mu^+$ or $\mu^-$. Thus from the nature 
of the track in ICAL,  the detection of a $\nu_{\mu}$ or a $\bar{\nu}_{\mu}$ 
can be confirmed.  

We consider such an ICAL detector at the proposed 
India-based Neutrino Observatory or INO \cite{ino}, which is at a baseline 
length $\sim$ 7150 Km from a proposed neutrino factory at CERN. 
This length, being very near to the magic baseline length
allows CP independent analysis of oscillation.
ICAL at INO is proposed to be a 50 KTon detector consisting of a stack of 
140 layers of iron plates of thickness 6 cm
and interleaved with 2.5 cm gap. This is proposed to have 
a dimension of 48 m $\times$ 16 m $\times$ 12 m when complete\cite{ino}.
The ICAL set up will be magnetised by a magnet of field strength $\sim 1.3$ 
Tesla. 

A flux of $\nu_{\mu}$ from a $\mu^-$ storage ring (from the decay 
$\mu^- \rightarrow e^- + \bar{\nu}_e + \nu_{\mu}$)
will suffer depletion on reaching the ICAL mentioned above 
due to oscillation of $\nu_\mu \rightarrow \nu_{\alpha_x}$, where 
$\alpha_x$ denotes any flavour other than $\mu$. This depleted $\nu_\mu$ 
flux will produce $\mu^-$ tracks at the magnetised ICAL
detector, in this disappearance channel. These $\mu^-$ are referred to as 
right sign muons.
If on the other hand ICAL registers $\mu^+$  
tracks from such a 
beam then this would imply the appearance of $\bar{\nu}_{\mu}$ at 
ICAL which can be obtained in the beam (of $\nu_\mu$ and $\bar {\nu}_e$)
due to the oscillation process 
$\bar{\nu}_e \rightarrow \bar{\nu}_{\mu}$. This is the appearance channel. 
Such $\mu^+$ are wrong sign muons. 

\begin{figure}[h!]
\centering
\subfigure[ Variation of $\Phi_{\nu_{\mu},\bar{\nu}_{\mu}}$ 
with $E_{\nu}$ at CERN-INO for injected muon energy
fixed at 50 GeV]{
\includegraphics[width=2.0in]{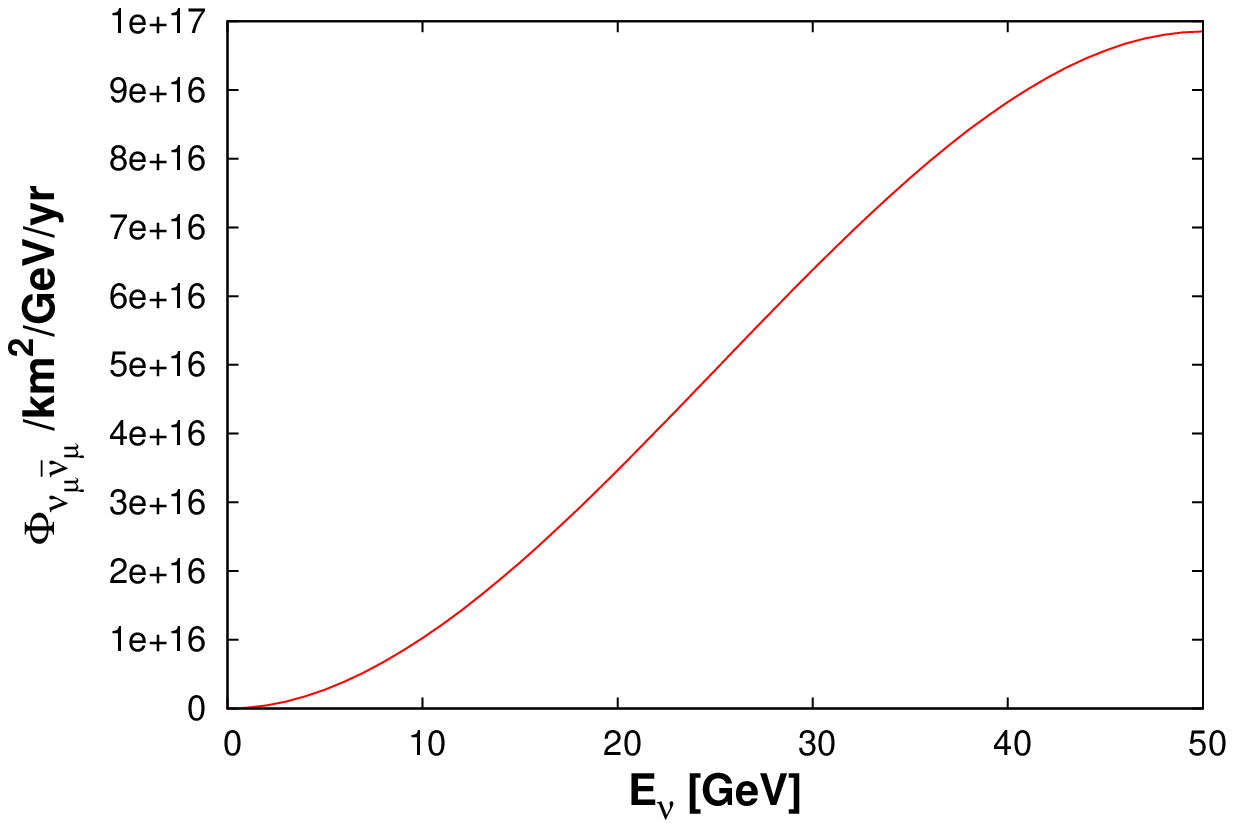}}
\subfigure[ Variation of $\Phi_{\nu_e,\bar{\nu}_e}$ with $E_{\nu}$ at 
CERN-INO for injected muon energy fixed
at 50 GeV]{
\includegraphics[width=2.0in]{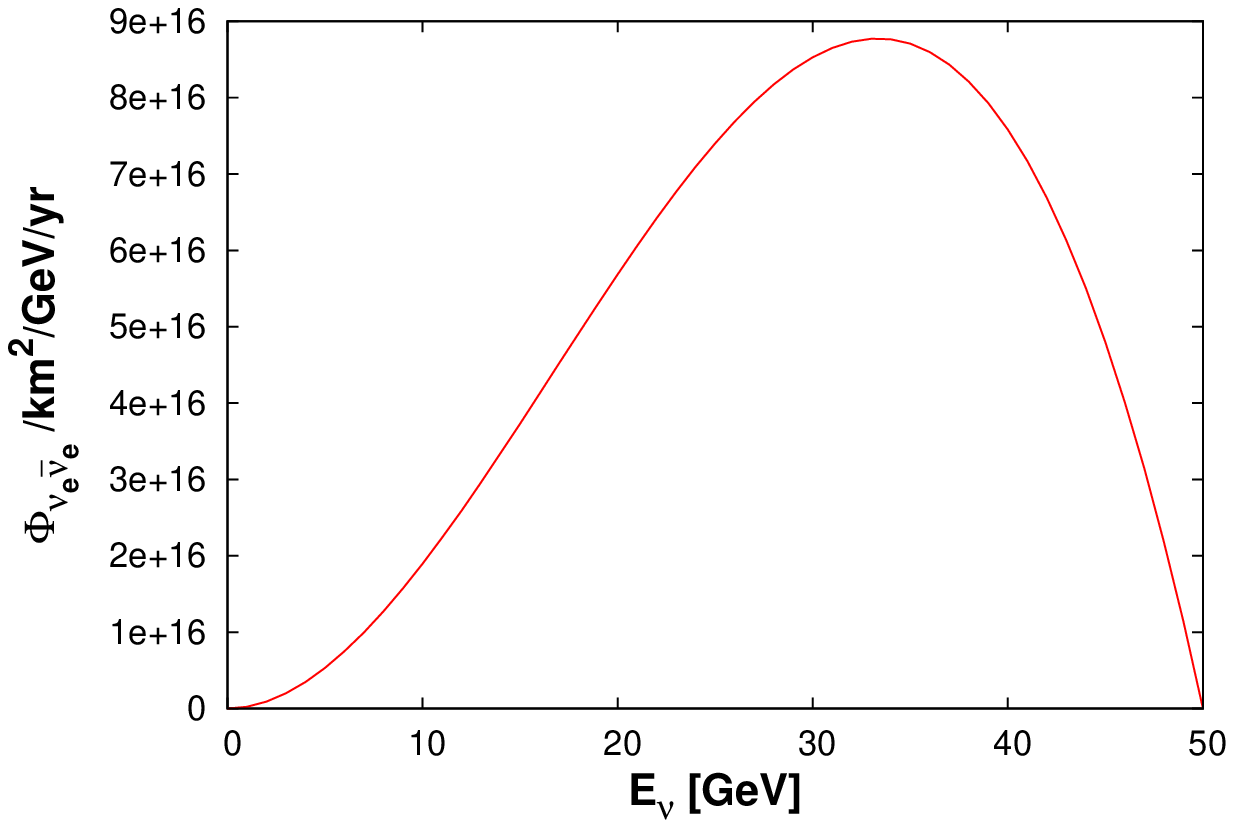}}
\caption{$\nu_{\mu}$($\bar{\nu}_{\mu}$) flux (plot a) and 
$\nu_e$($\bar{\nu}_e$) 
flux (plot b) for different neutrino energies ($E_{\nu}$) in case of 
a 50 GeV $\mu^-$ ($E_\mu = 50$ GeV) decaying 
in a storage ring for a period of 5 years. The baseline length 
$L = 7154.57$ Km  }
\label{fig4}
\end{figure}

In the present work, for the purpose of demonstration 
we have taken the injected muon energy at storage ring to 
be 50 GeV with the estimated baseline length of
CERN-ICAL distance. The disttance between CERN to the proposed 
site for the ICAL is calulated to be $L=7154.57$ Km. 
The average earth matter density ($\rho$) for this baseline is taken to be 
4.15 gm/cc.
The $\nu_{\mu}$ and $\bar{\nu}_e$ flux at the ICAL detector for such neutrino 
beam (assumed to have originated from CERN and contains in the beam
${\nu}_{\mu}$ and $\bar{\nu}_e$) 
are calculated using Eqs. \ref{eq29} - \ref{eq30} for 5 years of run and are 
plotted in plots (a) and (b) respectively of Fig. \ref{fig4}. 
\begin{figure}[ph]
\centerline{\includegraphics[width=2.0in]{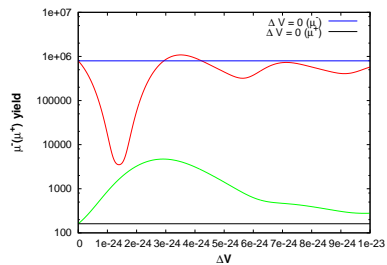}}
\vspace*{8pt}
\caption{Dependence of $\mu^{\pm}$ yield in the prescence of
velocity dependant oscillation of neutrinos\label{fig5}}
\end{figure} 
From the knowledge of the flux $\phi_{\nu_\mu}$ and $\phi_{\nu_e}$
(Fig.~\ref{fig4}), the oscillation probabilities 
$P_{\nu_\mu \rightarrow \nu_x}$ (for disappearance channel) 
and $P_{{\bar {\nu}_e} \rightarrow \bar {\nu}_{\mu}}$ (for appearance channel)
and the cross-section for the charged interactions of 
$\nu_\mu (\bar {\nu}_\mu)$ with iron that will produce $\mu^- (\mu^+)$
at ICAL, the yields of $\mu^-$ and $\mu^+$ can be computed.   
In this work we compute the $\mu^-$ (right sign) yields and $\mu^+$
(wrong sign) yields at the ICAL detector considered here
with oscillation probabilities that include velocity effects 
as well as matter effects for a baseline length of 7154.57 Km.
In Fig.~\ref{fig5} we demonstrate the variation of these $\mu$-yields as a
function of velocity difference ($\Delta V$) between two neutrino species 
while $\Delta V$ is varied from zero (no velocity oscillation) to 
$\Delta V = 10^{-23}$.
Red (green) line in Fig.~\ref{fig5} corresponds to the yield of right sign
(wrong sign) $\mu$ for a run period of 5 years.
The blue and black lines in Fig.~\ref{fig5} represent the right and wrong
sign muon yields respectively when $\Delta V = 0$. It is evident from
Fig.~\ref{fig5} that velocity induced oscillation significantly affects the
yields of $\mu$ when compared with the case of normal mass-flavour oscillation
($\Delta V=0$). It can also be observed from the plots in Fig.~\ref{fig5} that 
right sign $\mu$ yield decreaes rapidly for 
$\Delta V \sim 1-2\times10^{-24}$ and
then it saturates to a value appproximaltely to the yield when $\Delta V = 0$
for the remaining range ($\Delta V =10^{-23}$). Conversely yield of the wrong
sign $\mu$ increases nearly an
order of magnitude for $\Delta V \sim 1-5\times10^{-24}$. For higher values of
$\Delta V$, yields of wrong sign $\mu$ also tend to saturate to the yield
corresponding to normal mass flavour oscillation. One can also conclude
from Fig.~\ref{fig5} that considerable changes in both right and wrong sign 
$\mu$ yields are observed  near $\Delta V \sim 10^{-24}$.  
In Table 2 we tabulate the estimated yields 
for a run period of 5 years. 
Results are shown for the three values of velocity 
difference, namely, $\Delta V$ ($\Delta V = 10^{-23},~10^{-24},~10^{-25}$) 
as adopted in Section 2. The results with 
$\Delta V = 0$ (no velocity effect) are also given in Table 2 for comparison. 
\begin{table}[h]
\tbl{ The right sign $\mu$ yield and wrong 
sign $\mu$ yield at the ICAL detector with baseline length of 7154.57 Km in
case of different values of $\Delta V$ with injected muon energy fixed 
at 50 GeV.}
{\begin{tabular}{@{}c@{}} \toprule
~~~~~~~~~~~~~~~~~~~~~~~~~~CERN-ICAL ~~~~~~~~~~~~~~~~~~~~~~~~~~~~       \\ \botrule
\end{tabular} }
{\begin{tabular}{@{}c@{}}
~~~~~~~~~~~~~~~Baseline length ($L$) 7154.57 Km~~~~~~~~~~~~~~ \\  \botrule
\end{tabular}  }
{\begin{tabular}{@{}ccc@{}} 
$\Delta V $                        &  Right sign  $\mu$  & Wrong sign  $\mu$ \\ \botrule
~~~~~~~~~ 0.0        ~~~~~~~~~&   794147            &     161            \\
~~~~~~~~~ 1E-23      ~~~~~~~~~&   575795            &     278            \\
~~~~~~~~~ 1E-24      ~~~~~~~~~&   17132             &     871            \\ 
~~~~~~~~~ 1E-25      ~~~~~~~~~&  663325             &     183            \\ \botrule 
\end{tabular} \label{ta2} }
\end{table}

It is evident from the comparison of the calculated yields of right 
sign and wrong sign muons in Table~\ref{ta2} that the velocity 
induced oscillation indeed affects the yields significantly when compared
with the normal mass-flavour oscillation ($\Delta V =0$). Among
the 
chosen three values of $\Delta V$, while the 
effect of $\Delta V$ is negligible for $\Delta V = 10^{-25}$, 
this is most pronounced for the case when $\Delta V = 10^{-24}$. 
For the latter case, the calculated right sign muon is an order 
of magnitude less than for the case $\Delta V =0$ and 
there is almost an order of magnitude gain ($\sim 10^2$ for $\Delta V =0$
to $\sim 10^3$ with $\Delta V = 10^{-24}$)  for wrong sign 
muon when $\Delta V = 10^{-24}$
over the case with no velocity effect. More statistics for 
the wrong sign muons are in fact helpful for the study of 
crucial $\nu_e \rightarrow \nu_\mu$ oscillations. 

\section{Summary and Conclusion}
In the present work, the effect of velocity 
induced neutrino oscillation is explored in the 3-generation 
framework of neutrinos. This is based on the consideration that
the three different neutrinos can have three different 
maximum attainable velocities and their flavour 
eigenstates are not the same as their velocity eigenstates and mass 
eigenstates. The oscillation phenomena in this scenario are studied 
with modifying the Hamiltonian in the mass basis by adding the Hamiltonian in 
the velocity basis and then the evolution equations 
for the neutrino flavours are written down in terms of this modified 
Hamiltonian. Thus we have the formalism for neutrino flavour oscillations 
induced by both mass and velocity. Unlike for the case of purely mass-induced
flavour oscillation, where the phase difference of two neutrinos 
(that causes the oscillation) $\sim \frac {1} {E}$, for purely 
velocity-induced flavour oscillation the phase difference $\sim E$. 
Thus we study in this work, the combined effect of both of them 
on neutrino flavour oscillations in the realistic case of three 
generations. 

For demonstrative purposes, we consider the case of baseline neutrinos
with a representative baseline length of 7000 Km.
The oscillation probabilities 
are computed for this example first by considering no matter effect 
(vacuum oscillations) and then by incorporating the matter effect or 
MSW effect in the present formalism. Three values of $\Delta V$ 
($\Delta V = 10^{-23}, 10^{-24}, 10^{-25}$) are chosen for the calculation
and the results are compared with the case 
when no velocity effect is present ($\Delta V =0$). We found that 
near the vicinity of $\Delta V \sim 10^{-24}$ velocity induced oscillation
effect 
provides significant change in oscillation when comapred with
the normal mass flavour oscillation scenario for the case of long baseline
neutrino oscillation with baseline length $L \sim 7000$ Km. Also it 
appears that the choice of 
$\Delta V$ values, further order of magnitude lower than this value
(such as $\Delta V = 10^{-25}$), 
in fact produces almost or no velocity-induced effects on oscillation 
probabilities. On the other hand, when $\Delta V$ is chosen to be order 
of magnitude higher than the value of $10^{-24}$ (such as for 
$\Delta V = 10^{-23}$), the probabilities show rapid oscillatory nature.
It is also seen that while the MSW resonance condition is saisfied 
for two values of resonance energies for both the choices of
$\Delta V = 10^{-24}$ and $\Delta V = 10^{-25}$, the choice 
$\Delta V = 10^{-23}$ yields no resonance since such a choice does
not satisfy the resonance condition. The two values of resonance 
energies are however expected as here the effects of both
$\sim E$ and $\sim \frac {1} {E}$ dependences reduce the resonance
condition  to an equation quadratic in $E$.  

Hence the Hamiltonian of time evolution of neutrinos 
will be changed in presence of velocity mixing oscillation. 
We investigate the possible effects of the inclusion of the 
velocity dependent states that modify the normal mass oscillation 
scenario. In the present formalism, the unitary matrix connecting flavour and
mass basis also connects flavour basis to velocity basis.
This gives modified oscillation probability equations and we plot the 
different probabilities as a function of energy with and without 
presence of matter.
We study how the presence of velocity dependent states can 
change the different oscillation probabilities $P_{\alpha_1 \alpha_2}$ 
and its effects on resonance phenomena. We find that in some cases 
velocity induced oscillations differ significantly from normal (flavour - mass) oscillation of neutrinos.

We then apply our results for the case of a possible 
long baseline neutrino experiment. 
For this purpose we have considered a magnetised iron 
calorimeter detector (ICAL) at the proposed India-based 
Neutrino Observatory (INO) \cite{ino} as the end detector 
that would detect the muon neutrinos in a neutrino beam of 
$\nu_\mu$ and $\bar {\nu}_e$ considered 
to have originated from a possible neutrino factory at CERN having baseline
length $\sim 7154$ Km.
The muon neutrinos would produce muons following their charged current 
interactions with iron of ICAL and identifying the polarities of 
such muon signals, $\mu^-$ or $\mu^+$  
at ICAL, since the latter is magnetised, the nature of the 
detected neutrinos (whether it is $\nu_\mu$ or 
$\bar {\nu}_\mu$) can be concluded. 
While the $\mu^-$ events in the present example are termed as the right 
sign muons indicating the detection of $\nu_\mu$ the flux of which would
suffer depletion in the beam due to oscillation in disappearance channel, 
detecting of the wrong sign muons or $\mu^+$ 
definitely indicate the oscillation $\bar {\nu}_e \rightarrow \bar {\nu}_\mu$
in the appearance channel. In order to show how velocity induced oscillation
affects the $\mu^{\pm}$ yield, we plot $\mu^{\pm}$ yield as a function of
$\Delta V$ in the present formalism.
Here too we find that $\mu^{\pm}$ yield changes drastically for some choice of 
$\Delta V$ $(\sim 10^{-24}$). 
We also find that the estimated number of wrong sign muons
is considerably large in the range of $\Delta V \sim 1-5\times10^{-24}$ 
which is helpful for effective statistical analysis of this important 
disappearnce channel.  Finally it could be concluded that, neutrino oscillation
experiments in near future capable of probing the velocity differences between 
nuetrino species and also the respective $\mu$ yield are expected to provide
valuable information to enlighten the velocity induced neutrino
oscillation or rule out the possibility of such velocity induced neutrino
oscillation.

\end{document}